\documentclass[acmsmall,screen]{acmart}
\AtBeginDocument{%
  }

\acmJournal{TOSEM}

\usepackage{enumitem}
\usepackage{pifont}
\usepackage{multirow,makecell}
\usepackage{color,xcolor}
\usepackage{listings,amsfonts}
\usepackage{caption}
\usepackage{subcaption}
\usepackage{threeparttable}
\usepackage{bbding}
\usepackage{booktabs} 
\usepackage{longtable}
\usepackage{flushend}

\usepackage{xhfill}
\usepackage[skins]{tcolorbox}
\usepackage{varwidth}
\usepackage{graphicx,scalerel}
\usepackage{tabularx}

\makeatletter
\newtcolorbox{mybox}[1][]{
  enhanced,
  colframe=black, colback=white,
  sharp corners,
  boxrule=0.6pt,
  detach title,
  coltitle=black,
  colbacktitle=white,
  fonttitle=\bfseries,
  enlarge left by= 0mm,
  enlarge right by= 0mm,
  width=\linewidth,
  left=2mm,
  right=2mm,
  before upper=\setlength{\parindent}{17.62482pt}\everypar{{\setbox0\lastbox}\@minipagefalse\everypar{}},
  overlay={
    \node[anchor=east, 
          fill=tcbcolbacktitle, 
          font=\kvtcb@fonttitle]
          at ([xshift=95mm]frame.north west) % Posiziona il nodo
    {
      \begin{varwidth}{\linewidth}
        \centering\tcbtitle\par
      \end{varwidth}
    };
  },#1}
\makeatother

\AtEndPreamble{
	\usepackage{hyperref}

}

\begin{document}

\title{Large Language Model Supply Chain: A Research Agenda}

\author[S Wang]{Shenao Wang}
\authornote{Hubei Key Laboratory of Distributed System Security, Hubei Engineering Research Center on Big Data Security, School of Cyber Science and Engineering, Huazhong University of Science and Technology.}
\email{shenaowang@hust.edu.cn}
\orcid{0000-0003-3818-3343}
\affiliation{%
  \institution{Huazhong University of Science and Technology}
  \city{Wuhan}           
  \country{China}
}

\author[Y Zhao]{Yanjie Zhao}
\authornotemark[1]
\email{yanjie_zhao@hust.edu.cn}
\orcid{0000-0001-8793-5367}
\affiliation{%
  \institution{Huazhong University of Science and Technology}
  \city{Wuhan}           
  \country{China}
}

\author[X Hou]{Xinyi Hou}
\authornotemark[1]
\email{xinyihou@hust.edu.cn}
\orcid{0009-0005-9965-2109}
\affiliation{%
  \institution{Huazhong University of Science and Technology}
  \city{Wuhan}           
  \country{China}
}

\author[H Wang]{Haoyu Wang}
\authornote{Haoyu Wang is the corresponding author (haoyuwang@hust.edu.cn).}
\authornotemark[1]
\email{haoyuwang@hust.edu.cn}
\orcid{0000-0003-1100-8633}
\affiliation{%
  \institution{Huazhong University of Science and Technology}
  \city{Wuhan}           
  \country{China}
}

\renewcommand{\shortauthors}{Shenao Wang, Yanjie Zhao, Xinyi Hou, \& Haoyu Wang}

\begin{abstract}
The rapid advancement of large language models (LLMs) has revolutionized artificial intelligence, introducing unprecedented capabilities in natural language processing and multimodal content generation. However, the increasing complexity and scale of these models have given rise to a multifaceted supply chain that presents unique challenges across infrastructure, foundation models, and downstream applications. This paper provides the first comprehensive research agenda of the LLM supply chain, offering a structured approach to identify critical challenges and opportunities through the dual lenses of software engineering (SE) and security \& privacy (S\&P). We begin by establishing a clear definition of the LLM supply chain, encompassing its components and dependencies. We then analyze each layer of the supply chain, presenting a vision for robust and secure LLM development, reviewing the current state of practices and technologies, and identifying key challenges and research opportunities. This work aims to bridge the existing research gap in systematically understanding the multifaceted issues within the LLM supply chain, offering valuable insights to guide future efforts in this rapidly evolving domain.
\end{abstract}

\begin{CCSXML}
<ccs2012>
   <concept>
       <concept_id>10011007.10011074.10011134.10003559</concept_id>
       <concept_desc>Software and its engineering~Open source model</concept_desc>
       <concept_significance>500</concept_significance>
       </concept>
   <concept>
       <concept_id>10011007.10011006.10011072</concept_id>
       <concept_desc>Software and its engineering~Software libraries and repositories</concept_desc>
       <concept_significance>500</concept_significance>
       </concept>
   <concept>
       <concept_id>10002944.10011122.10002945</concept_id>
       <concept_desc>General and reference~Surveys and overviews</concept_desc>
       <concept_significance>500</concept_significance>
       </concept>
   <concept>
       <concept_id>10010147.10010178</concept_id>
       <concept_desc>Computing methodologies~Artificial intelligence</concept_desc>
       <concept_significance>500</concept_significance>
       </concept>
 </ccs2012>
\end{CCSXML}

\ccsdesc[500]{Software and its engineering~Open source model}
\ccsdesc[500]{Software and its engineering~Software libraries and repositories}
\ccsdesc[500]{General and reference~Surveys and overviews}
\ccsdesc[500]{Computing methodologies~Artificial intelligence}

\keywords{LLM Supply Chain, Large Language Models}

\maketitle

\section{Introduction}
The landscape of artificial intelligence (AI) has been dramatically transformed by the rapid advancements in pre-trained large language models (LLMs) and multimodal large language models (MLLMs)~\footnote{For simplicity in this text, both pre-trained LLMs and MLLMs will be collectively referred to as LLMs, and their supply chains will be referred to as the LLM supply chain in the subsequent sections.}. Models such as GPT-4~\cite{achiam2023gpt}, Gemini~\cite{team2023gemini}, and LLaMA~\cite{touvron2023llama} have pushed the boundaries of what machines can understand and generate, exhibiting remarkable capabilities in processing and producing human-like text and multimodal content. These sophisticated models are trained on vast and diverse datasets, enabling them to perform a wide array of tasks ranging from natural language understanding and content generation. 

As the adoption of LLMs continues to expand, the necessity for a robust and efficient supply chain to support their development, deployment, and maintenance becomes increasingly apparent. The LLM supply chain encompasses the lifecycle of these models, involving multiple stages such as data curation~(e.g. clearnlab~\cite{cleanlab} for data quality assurance and Hugging Face datasets~\cite{datasets} for data management), model training~(e.g. PyTorch Distributed~\cite{pytorch} for distributed training), optimization~(e.g. OmniQuant~\cite{OmniQuant} for model quantization and mergekit~\cite{mergekit} for model merging), and deployment~~(e.g. AutoGPT~\cite{AutoGPT} for agent workflow orchestration and RAGFlow~\cite{RAGFlow} for retrieval-augmented generation). This complex ecosystem requires seamless collaboration among various stakeholders, including model developers who design and refine architectures, data providers who supply necessary training data, infrastructure engineers who manage computational resources, and end-users who interact with final applications. Each stakeholder plays a critical role in ensuring the functionality and reliability of LLMs, and the interdependencies among them highlight the need for coordinated efforts to address challenges inherent in the supply chain.

\noindent \textbf{LLM Supply Chain.} Central to the LLM supply chain are three layers that collectively determine the effectiveness and efficiency of deploying LLMs in real-world applications. First, the \textit{LLM infrastructure layer} comprises the computational resources, datasets, and toolchains, which are essential for training, optimizing, and deploying the models. This includes not only the vast amounts of data required to train LLMs but also the software frameworks and hardware resources necessary to handle such scale. Second, the \textit{foundation model layer} covers the sequential processes of training, testing, releasing, and ongoing maintenance. Each phase demands meticulous attention to detail to ensure that the models perform as intended and can adapt to new information or requirements over time. Third, the \textit{downstream application ecosystem} enables the integration of pre-trained models into diverse intelligent applications, allowing businesses and users to harness LLM capabilities and translate their potential into tangible benefits across various domains.

\noindent \textbf{Current State \& Challenges.} Despite the significant progress and transformative potential of LLMs, the rapidly evolving landscape of their development and deployment presents many challenges across the entire supply chain. These challenges span multiple fileds, including security, privacy, software engineering, systems, networking, data science, and user experience design. In this paper, we focus specifically on \textbf{\underline{Software Engineering~(SE)}} and \textbf{\underline{Security \& Privacy~(S\&P)}}.
From the SE perspective, the unique characteristics of LLMs introduce a host of challenges that extend beyond traditional software development practices. The sheer scale and complexity of these models necessitate novel approaches to system design, development, and maintenance. Key challenges include model versioning and reproducibility~\cite{ajibode2024towards}, dependency management~\cite{jiang2023empirical}, adapting continuous integration and deployment (CI/CD) practices for LLMs~\cite{chen2023llmops}, comprehensive testing and quality assurance~\cite{chang2024survey,sun2024trustllm}, model interpretability and explainability~\cite{singh2024rethinking}, technical debt management~\cite{bill2024llmdebt}, and navigating the complex landscape of open-source licenses~\cite{charlie2023llamalicense,sriram2024llama3license}.
From the S\&P perspective, vulnerabilities have been identified at various layers of the LLM supply chain. These range from flaws in LLM development toolchains at the infrastructure layer~\cite{mlflow2024cve,hiddenlayer2023keras,peng2024gguf,smartkeyss2024nltkrce} to vulnerabilities~\cite{eoin2024hfconversion} or code poisoning~\cite{zhao2024models} attacks in model hubs, and extend to remote code execution risks in downstream applications~\cite{liu2023demystifying}. The pervasive nature of these vulnerabilities highlights the need for comprehensive security measures throughout the LLM supply chain.
The potential for cascading effects due to model reuse and the difficulty in isolating and mitigating issues in complex LLM-based systems further exacerbate these challenges, often outstripping the establishment of robust security measures and software engineering practices.

\noindent \textbf{Research Gap \& Our Work.} Addressing these challenges necessitates a multidisciplinary approach that draws insights from fields such as software engineering, system architecture, security, and data governance. However, there exists a significant research gap in systematically tackling these multifaceted issues within the context of the LLM supply chain. 
To bridge this gap, \underline{\textbf{we present the first comprehensive definition of the LLM supply chain}}, proposing a structured approach to analyze its challenges and opportunities across three key layers: infrastructure, foundation models, and downstream applications. We begin by establishing a clear definition of the LLM supply chain, encompassing its components and dependencies. Building upon this foundation, we examine each layer through the dual lenses of SE and S\&P. For each layer, we present a vision for the robust and secure LLM supply chain, review the current state of practices and technologies, and identify key challenges and research opportunities. In the infrastructure layer, we explore issues related to computational resources, development tools, and datasets. For foundation models, we delve into the complexities of model training, testing, sharing, and maintenance. The downstream applications layer focuses on the integration and deployment of LLM-based systems in real-world scenarios. Through this comprehensive examination, we propose a research agenda that not only identifies current challenges but also anticipates future developments in the field. Our goal is to offer insights that can guide future efforts in this rapidly evolving domain, fostering innovation while upholding ethical standards and ensuring the responsible advancement of LLM technologies.

In summary, our contributions are detailed as follows:

\begin{itemize}[leftmargin=15pt]

\item \textbf{Definition of the LLM supply chain}: We are the first to present the definition of the LLM supply chain, which provides a foundational understanding of the LLM supply chain's components and stages. The supply chain is delineated into three core layers: the model infrastructure layer (including computational resources, datasets, and toolchains for training, optimization, and deployment), the foundation model layer (encompassing training, testing, releasing, and ongoing maintenance), and the downstream application ecosystem (facilitating the integration of pre-trained models into various intelligent applications).

\item \textbf{Identification of Critical Challenges}: For each layer of the LLM supply chain, we offer a vision for robust and secure LLM development and deployment, review the current state of practices and technologies, and identify key challenges, all through the dual lenses of SE and S\&P. The SE perspective focuses on system design, development processes, quality assurance, and license compatibility, while the S\&P viewpoint addresses vulnerabilities, privacy protection, and bias mitigation, providing a comprehensive analysis that spans the entire supply chain.

\item \textbf{Proposal of the Research Agenda}: Based on the systematic review, we propose a forward-looking research agenda that addresses current challenges and anticipates future developments in the LLM supply chain. This agenda is designed to guide future efforts in the rapidly evolving domain of the LLM supply chain by suggesting multidisciplinary approaches that draw from SE and S\&P. The goal is to foster advancements that maximize the transformative potential of LLMs while adhering to ethical standards and promoting sustainability. 

\end{itemize}
\section{Definitions and Significance}
In this section, we start by defining the LLM supply chain (\autoref{sec:definition}). We then explore three key aspects that highlight its significance: the novel challenges introduced by LLMs beyond traditional open-source software (OSS) (\autoref{sec:oss}), the increased complexity of managing LLMs compared to conventional machine learning (ML) and deep learning (DL) models (\autoref{sec:mldl}), and the cascading implications beyond individual models or applications (\autoref{sec:Responsible}). \textbf{These elements collectively underscore the urgent need for a comprehensive research agenda focused on the LLM supply chain.}

\begin{figure}[t]
    \centering
    \includegraphics[width=0.99\linewidth]{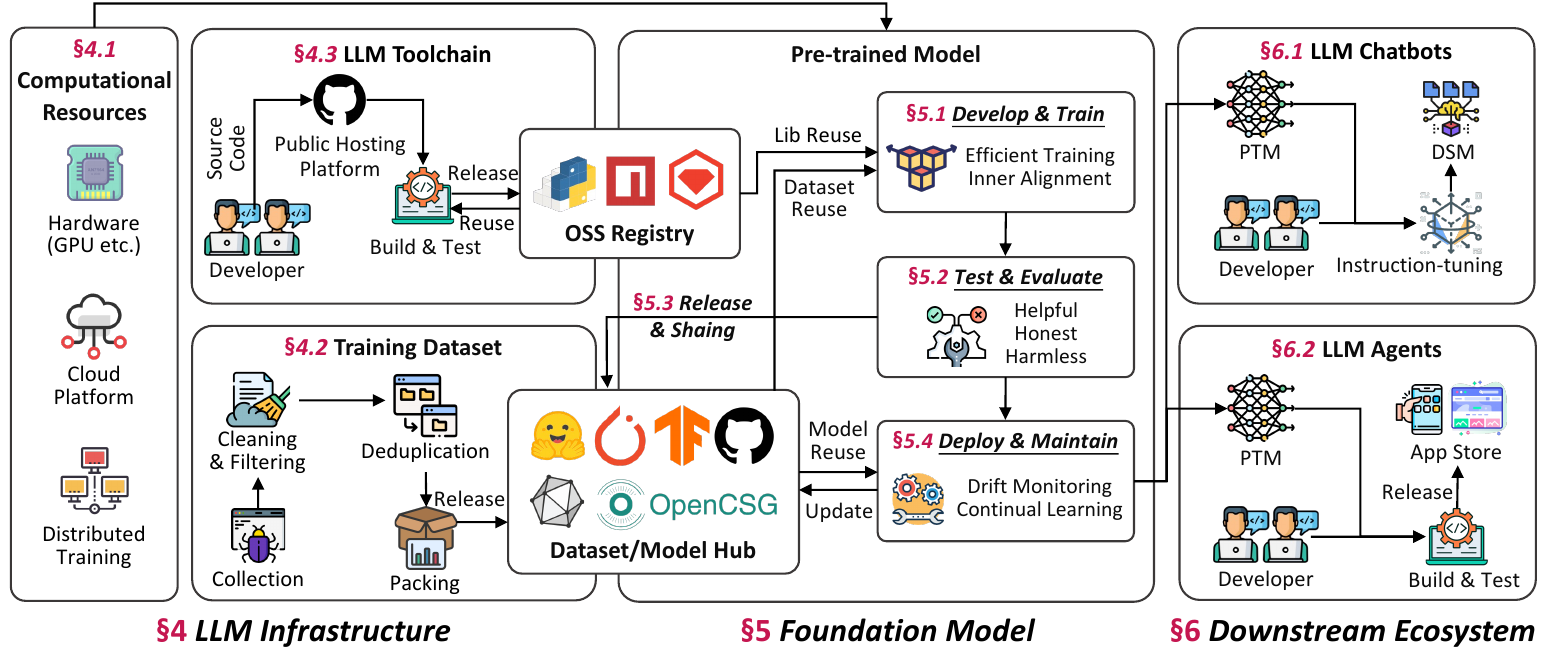}
    \caption{Definition and Each Component of the LLM Supply Chain.}
    \label{fig:structure}
\end{figure}

\subsection{Definitions}
\label{sec:definition}
In the traditional domain of the OSS supply chain, numerous efforts~\cite{ji2023oss,gao2024oss,he2020oss} have been made to define the scope and components of OSS supply chains. He et al.~\cite{he2020oss} defined the OSS supply chain as the end-to-end process of developing, releasing, deploying, and maintaining software from source code to final delivery to users. Ji et al.~\cite{ji2023oss} adopted a broader perspective, incorporating third-party components and defining the OSS supply chain as the network of supply relationships formed by upstream communities, source/binary packages, distribution markets, developers, maintainers, and foundations during software development and operation. Gao et al.~\cite{gao2024oss} further delineated four types of open-source artifact reuse relationships~(dependency installation, API calls, file copying, and code cloning), characterizing the OSS supply chain as the supply network arising from various code reuse practices across the software lifecycles.

Building upon these prior conceptualizations of OSS supply chains, we define the LLM supply chain as a complex network system $S = (C, D)$, where $C$ represents the set of components and $D$ represents the set of dependencies. This system involves all upstream data providers, model development communities, model repositories, distribution platforms, and app markets, as well as data scientists, researchers, engineers, maintainers, and end-users in the process of developing, distributing, and deploying pre-trained models. As illustrated in \autoref{fig:structure}, the LLM supply chain comprises five key components $C = \{R, T, D, M, A\}$: computational resources ($R$), tools ($T$), datasets ($D$), models ($M$), and applications ($A$). The first three components ($R$, $T$, and $D$) collectively form the infrastructure of the LLM supply chain.
\begin{itemize}[leftmargin=15pt]
    \item Computational Resources ($R$): Includes high-performance computing clusters, GPUs, TPUs, cloud computing platforms, and distributed computing systems.
    \item LLM Toolchain ($T$): Encompasses LLM development toolchains, including AI frameworks~(e.g. PyTorch~\cite{pytorch} and TensorFlow~\cite{tensorflow}), dataset curation tools, model optimization tools~(e.g. OmniQuant~\cite{OmniQuant} for model quantization and mergekit~\cite{mergekit} for model merging), and downstream application frameworks~(e.g. LangChain~\cite{langchain}).
    \item Datasets ($D$): Comprises large-scale text corpora~(including natural language and code), specialized domain datasets, and multi-modal data collections.
    \item Models ($M$): Represents the various stages of LLM development, including architecture design, pre-training, fine-tuning, testing, releasing, deployment, and maintenance.
    \item Applications ($A$): Includes downstream LLM-based applications such as chatbots, autonomous agents, and domain-specific LLM solutions.
\end{itemize}

Within the LLM supply chain, there exist multiple complex dependencies arising from reuse and sharing of components across different stages. These dependencies can be denoted as $D = \{d_{ij} | i,j \in C\}$, where $C$ represents the set of components in the supply chain. Specifically, these dependencies are primarily driven by reuse and can be summarized as follows:

\begin{itemize}[leftmargin=15pt]
\item $d_{RM}$: Vulnerabilities in computational resources may lead to single points of failure, impacting model availability and security.
\item $d_{TT}$: Interoperability and version compatibility issues between different development tools may introduce vulnerabilities or security flaws in the toolchain.
\item $d_{TD}$: Deficiencies in data processing tools can compromise dataset quality and integrity, affecting the security and fairness of models trained on these datasets.
\item $d_{TM}$: Reused tools and components in the development process may introduce vulnerabilities or unintended optimizations into models and applications.
\item $d_{TA}$: The security and functionality of development tools can directly impact the security and reliability of LLM applications built using these tools.
\item $d_{DD}$: Interconnections and shared usage between datasets, either through direct reuse or derivation from common sources, can lead to the propagation of data quality issues. 
\item $d_{DM}$: Data quality issues can severely impact the performance, security, and fairness of models trained on these datasets.
\item $d_{MM}$: Dependencies between models, such as transfer learning or model merging, may facilitate the propagation of vulnerabilities or biases across different models.
\item $d_{MA}$: Applications may inherit characteristics, vulnerabilities, and biases from the models they employ, affecting their security and fairness performance.
\end{itemize}

These dependencies not only form the core structure of the LLM supply chain but also serve as potential sources of security issues. Consequently, the scope of LLM supply chain security has expanded from individual models to encompass every component and dependency within the entire chain. This expansion significantly increases potential security risks, as the attack surface is no longer limited to a single model but covers the entire supply chain, including computational infrastructure, development tools, training data, and upstream models.
More importantly, due to the interdependencies within the supply chain, security issues in any component or dependency can trigger a chain reaction, affecting all downstream elements and thus amplifying the impact of security incidents. For example, a vulnerability in computational resources could compromise the integrity of model training~($d_{RM}$), leading to widespread model vulnerabilities that propagate to numerous applications~($d_{MA}$). Similarly, biases in a fundamental dataset could permeate through various models~($d_{DM}$) and ultimately affect the fairness of a wide range of applications~($d_{MA}$).
This complex dependency network underscores the importance of comprehensively considering supply chain security and provides a more systematic and holistic perspective for researching and addressing related security issues. Future research should focus on developing methodologies for tracking multi-level dependency relationships, assessing cascading impacts of vulnerabilities across different components, and implementing security measures that consider the entire LLM supply chain ecosystem, from computational infrastructure to end-user applications.

\begin{table}[t]
\centering
\fontsize{8}{11}\selectfont
\caption{Comparison of the Traditional OSS Supply Chain and the LLM Supply Chains.}
\label{tab:supply-chain-comparison}
\begin{tabular}{|p{2.5cm}|p{5cm}|p{5.5cm}|}
\hline
\textbf{Component} & \textbf{OSS Supply Chain} & \textbf{LLM Supply Chain} \\
\hline
Resources & Source code, documentation, repository commits \& license, package metadata, CI/CD configurations, etc. & Large-scale datasets, pre-trained models, prompts, documentation, repository license, model cards, etc. \\
\hline
Dev Tools & Version control systems (git etc.), third-party packages/libraries, IDEs, compilers, debuggers, build systems, static analysis tools, etc. & AI foundational frameworks (PyTorch, TensorFlow, etc.), third-party components \& tools (RAG, vector database, quantization, etc.), model hubs~(Hugging Face, etc.), etc. \\
% \hline
% Core Product & Software applications, third-party packages/libraries, APIs, etc. & Pre-trained models, datasets, prompts, chatbots, agents, etc. \\
\hline
Distribution & Code repositories (GitHub, GitLab, etc.), package managers (NPM, PyPI, etc.), online platforms ~(Stack Overflow etc.) & Model hubs (Hugging Face, ModelScope, etc.), dataset communities~(Kaggle etc.), prompt hubs, code repositories, etc. \\
% \hline
% Deployment & Software installation, cloud deployment, containerization (Docker) & Model serving, API integration, containerized model deployment \\
\hline
Maintenance & Feature updates, bug fixes, vulnerability patches, documentation updates, etc. & Model retraining, fine-tuning, model drift monitoring, model merging, etc. \\
\hline
Quality Assurance & Unit testing, code reviews, specification checks, fuzz testing, SAST, IAST, etc. & Performance evaluation, bias \& privacy assessment, AI red teaming, etc. \\
\hline
\hline
\multicolumn{3}{|c|}{\textbf{Research Topics: SE Perspective}} \\
\hline
\textbf{Inherited Issues} & \multicolumn{2}{p{10.5cm}|}{Code clone (source code, binary), third-party library dependency, open-source licenses (conflicts \& compliance), repository measurement, code quality, etc.} \\
\hline
\textbf{New in LLMs} & \multicolumn{2}{p{10.5cm}|}{Model clone detection, model dependency management, risk propagation (hallucination, bias \& toxicity), model license compliance, prompt engineering best practices, etc.} \\
\hline
\multicolumn{3}{|c|}{\textbf{Research Topics: S\&P Perspective}} \\
\hline
\textbf{Inherited Issues} & \multicolumn{2}{p{10.5cm}|}{Source code vulnerabilities, code dependency \& vulnerability propagation, patch availability, supply chain poisoning attacks, SBOM, etc.} \\
\hline
\textbf{New in LLMs} & \multicolumn{2}{p{10.5cm}|}{Dataset poisoning, model deserialization attacks (e.g., backdoors), code hallucination, prompt injection, jailbreak attacks, vulnerable autonomous agents, etc.} \\
\hline

\hline
\end{tabular}
\end{table}

\subsection{Beyond OSS Supply Chain: The New Frontier}
\label{sec:oss}

While LLM and OSS supply chains share some common elements, they exhibit significant differences in their composition and characteristics. \autoref{tab:supply-chain-comparison} provides a comparative analysis of these two supply chains across various components.

\noindent \textbf{Compositional Differences.} The compositional differences between the LLM supply chain and the OSS supply chain contribute to the divergence in their supply chains, resulting in the LLM supply chain inheriting some research topics from the OSS supply chain while also introducing novel challenges. In terms of resources, OSS primarily relies on source code, documentation, repository commits, and package metadata, whereas LLM supply chains are built upon large-scale datasets, pre-trained models, prompts, and model-specific metadata like model cards. Besides, the OSS utilizes traditional software development tools, while LLM development necessitates AI-specific frameworks, specialized hardware, and tools for advanced tasks like retrieval-augmented generation (RAG), model quantization, and model merging. The distribution channels of these supply chains also differ significantly: OSS primarily leverages code repositories and package managers, in contrast to LLM products often being distributed through model hubs and AIOps platforms.  Similarly, OSS maintenance centers on feature updates and bug fixes, while LLM maintenance encompasses model fine-tuning, retraining, and monitoring for model drift. Lastly, quality assurance methods diverge, with OSS relying on traditional software testing approaches like unit testing and code reviews, in contrast to LLM quality assurance, which involves model-specific evaluations such as bias assessment and AI red teaming.

\noindent \textbf{Research Topics.} It's important to note that the LLM supply chain, in many ways, encompasses the OSS supply chain. The development components and tools used in LLM pipelines are often open-source software themselves. Many research questions in the LLM supply chain can be derived from or inspired by existing OSS supply chain research problems. However, the unique characteristics of LLMs also give rise to novel research topics that extend beyond traditional OSS concerns. 

\begin{itemize}[leftmargin=15pt]
    \item \textbf{SE Perspective:} From the SE standpoint, LLM supply chains inherit issues related to code clones, third-party library dependencies, and open-source license compliance. However, they also introduce new challenges specific to pre-trained models, which include model clone detection, model dependency management and model license compliance.
    
    \item \textbf{S\&P Perspective:} LLM supply chains inherit several security concerns from OSS, including source code vulnerabilities, dependency and vulnerability propagation, and supply chain poisoning attacks. However, LLMs introduce novel security challenges such as dataset poisoning, model deserialization attacks (e.g., backdoors), and code hallucination. Additionally, LLMs raise concerns about prompt injection, jailbreak attacks, and vulnerable autonomous agents.
\end{itemize}

% In conclusion, while the LLM supply chain shares some commonalities with the OSS supply chain, it presents a unique set of challenges and research opportunities. The intertwined nature of OSS and LLM supply chains necessitates an interdisciplinary approach to research, combining insights from responsible AI, security, and software engineering, to address both inherited and novel challenges in this evolving landscape.

\subsection{Beyond ML/DL Supply Chain: Scaling Complexity}
\label{sec:mldl}
The LLM supply chain represents a significant leap in complexity and scale compared to traditional ML and DL supply chains. ML/DL supply chains primarily focus on model development, training, and deployment for specific tasks, whereas LLM supply chains encompass a broader ecosystem with intricate interdependencies and far-reaching implications across multiple domains. This increased complexity introduces novel challenges and research opportunities.

\noindent \textbf{Infrastructure Layer.} At the infrastructure layer, LLMs demand unprecedented computational resources, necessitating the extensive use of distributed training frameworks, which introduces new security risks and challenges~\cite{valdez2024rpc} that were less prevalent in traditional ML/DL pipelines. The LLM era has also spawned a plethora of new development tools and features, from vector retrieval systems~\cite{han2023comprehensive,facebookresearch2024faiss} to inference optimization~\cite{ggerganov2024ggml} and deployment platforms~\cite{mlc2024mlcllm}. The security characteristics and software engineering best practices for these rapidly evolving toolchains remain understudied, potentially propagating vulnerabilities throughout the LLM supply chain. Moreover, datasets in the LLM era have gained unprecedented importance, raising critical issues such as privacy concerns~\cite{carlini2021extracting,kim2024propile}, bias mitigation~\cite{ferrara2023should,kirk2021bias}, and license compatibility~\cite{karamolegkou2023copyright,katzy2024codellmlicense} - topics that were seldom discussed in the ML/DL era.

\noindent \textbf{Foundation Model Layer.} In the foundation model layer, the reuse of pre-trained LLMs is far more common and consequential than in traditional ML/DL approaches. This shift introduces a critical challenge: the potential for cascading effects, where issues such as privacy violations~\cite{carlini2021extracting,lukas2023analyzing}, biases~\cite{gallegos2023bias,kirk2021bias}, or security vulnerabilities~\cite{yi2024jailbreak,zhou2024easyjailbreak} can propagate from foundation models to a multitude of downstream models and applications. The severity and breadth of this impact far exceed what was typically encountered in ML/DL supply chains, necessitating the need for more rigorous and comprehensive testing and evaluation protocols. Comprehensive testing now includes functionality and performance evaluation~\cite{dong2023codescore,sun2024scieval}, assessing the model's capabilities across a wide range of tasks and domains; reliability and truthfulness evaluation~\cite{sun2024trustllm,li2023halueval,lin2021truthfulqa}, measuring the model's consistency and tendency to generate factual information; and safety and ethical compliance evaluation~\cite{chao2024jailbreakbench,gehman2020realtoxicityprompts,huang2023flames}, examining the model's adherence to ethical guidelines and safety constraints. Furthermore, the proliferation of model hubs for sharing and distributing pre-trained models introduces new security considerations. The increased reliance on shared models in the LLM era has led to growing scrutiny of model hub security. Potential vulnerabilities in these platforms could lead to model or dataset poisoning~\cite{zhao2024models}, compromising model integrity on a scale unseen in the ML/DL era.

\noindent \textbf{Downstream Application Ecosystems.} The downstream application layer for LLMs introduces unprecedented complexity. LLMs are being integrated into a wide array of novel application forms, such as chatbots and LLM agents, which are significantly more sophisticated than traditional ML/DL applications. For instance, LLM agents often incorporate components like retrieval-augmented generation (RAG) systems and code execution sandboxes~\cite{li2024personal,he2024emerged}, expanding the attack surface and introducing security challenges that were non-existent in the ML/DL era~\cite{owasp2024llmv2}. LLM-powered chatbots face new threat models that are not relevant in traditional ML/DL applications. These include prompt injection attacks~\cite{liu2023prompt,greshake2023not}, where malicious users attempt to manipulate the model's behavior through carefully crafted inputs; function calling attacks~\cite{owasp2024llmv2}, exploiting the model's ability to call external functions, potentially leading to unauthorized actions; and prompt leakage~\cite{su2024gpt,hou2024security}, the unintended disclosure of sensitive information embedded in prompts or model responses.
As more sophisticated LLM agents emerge, additional concerns arise, such as agent autonomy escalation~\cite{owasp2024llmv2}, where agents potentially exceed their intended level of autonomy, leading to unpredictable or undesired behaviors. These challenges highlight the need for robust safety measures, continuous monitoring, and ethical guidelines in LLM-powered applications, considerations that were less critical in traditional ML/DL deployments.

In conclusion, while LLM supply chains build upon the foundations laid by ML/DL supply chains, they present a unique set of challenges and research opportunities that span the entire lifecycle of LLMs. The unprecedented scale and complexity of LLMs necessitate interdisciplinary research approaches to address both inherited and novel challenges in this rapidly evolving landscape.

\subsection{Beyond Responsible/Trustworthy AI: Cascading Impacts}
\label{sec:Responsible}

The emergence of LLMs has necessitated a paradigm shift in how we approach AI development, testing, and deployment. While the concepts of responsible/trustworthy AI and the LLM supply chain both address critical aspects of AI systems, they differ significantly in their focus, scope, and research priorities. These differences highlight the evolving nature of AI research and the need for more comprehensive frameworks to address the unique challenges posed by LLMs.

\noindent \textbf{Concept \& Research Scope.} Responsible/trustworthy AI has traditionally focused on ethical considerations, fairness, transparency, and accountability in AI systems~\cite{kenthapadi2023responsibleai,sun2024trustllm,kaur2022trustworthy}, primarily concerning itself with the societal impact and ethical implications of individual models or applications. In contrast, the LLM supply chain perspective, as explored in this paper, encompasses a broader view that includes both S\&P and SE aspects, which can be seen as a superset of responsible/trustworthy AI concepts.
While responsible/trustworthy AI typically concentrates on the properties of individual models~\cite{sun2024trustllm}, the LLM supply chain framework examines the cascading effects between models and throughout the entire LLM ecosystem. It considers how vulnerabilities, biases, or other issues can propagate through interconnected components, affecting multiple models and applications downstream. 
Furthermore, the LLM supply chain perspective incorporates aspects that are often overlooked in responsible/trustworthy AI discussions, such as the impact of infrastructure security on downstream models and applications. This broader view considers the entire technological stack supporting LLMs, from hardware to software infrastructure, and how vulnerabilities at any level can affect the entire ecosystem. For instance, one vulnerability in the infrastructure could potentially compromise multiple models and applications simultaneously~\cite{rayattacks,nelson2024llmserver}, leading to data leaks, model theft, or unauthorized access. Traditional responsible AI frameworks may not fully address these infrastructure-level risks and their far-reaching consequences. 
By encompassing both the foundational aspects of responsible AI and the wider considerations of system-level interactions and security, the LLM supply chain provides a more holistic perspective to understanding and mitigating risks in large-scale LLM systems. 
\section{Research Agenda Overview}

\begin{figure}[t]
    \centering
    \includegraphics[width=0.95\linewidth]{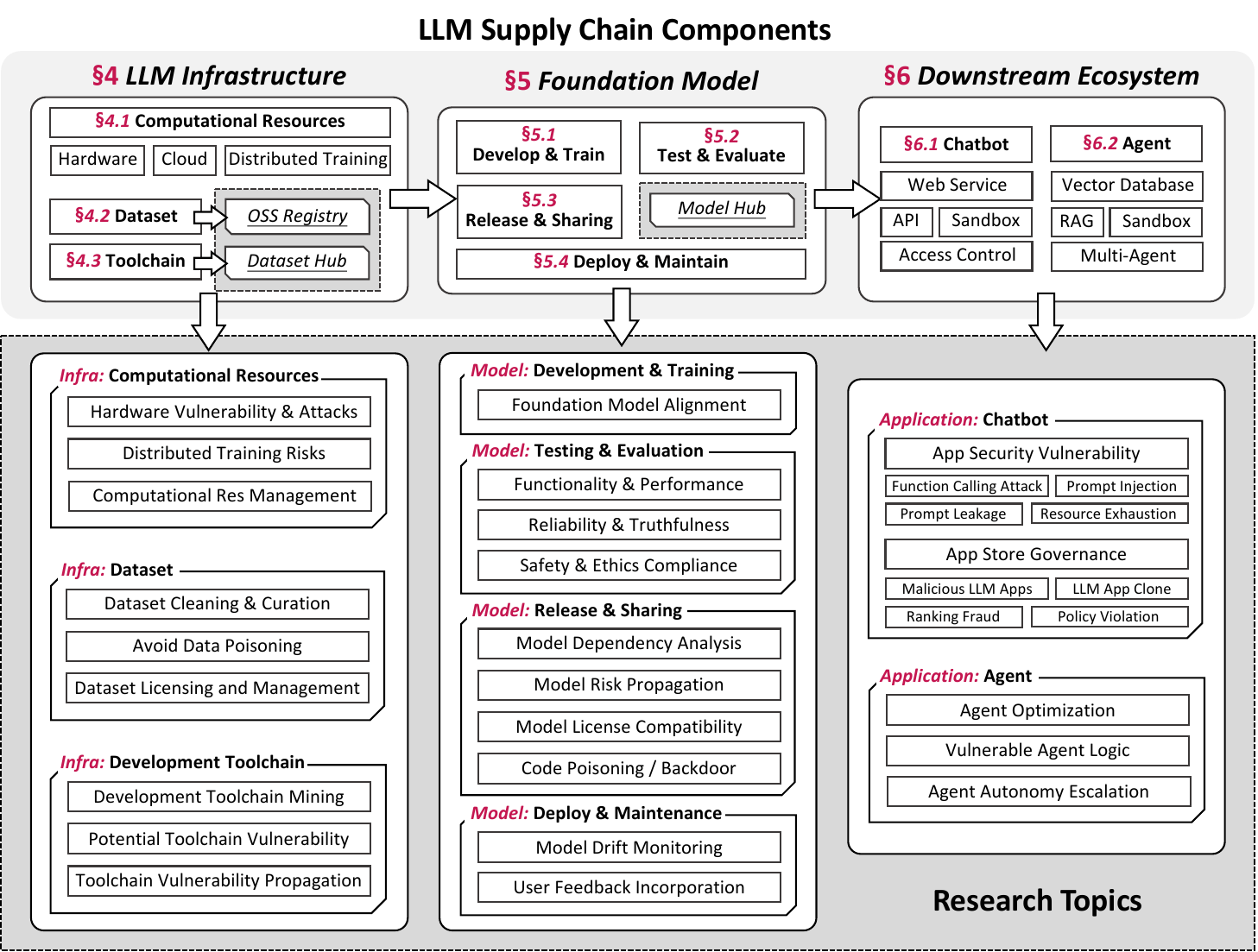}
    \caption{Research Agenda for the LLM Supply Chain.}
    \label{fig:agenda}
\end{figure}

The LLM supply chain encompasses a complex ecosystem of components, as illustrated in \autoref{fig:agenda}, each presenting unique challenges and opportunities for research from both \textbf{SE} and \textbf{S\&P} perspectives. This research agenda can be broadly categorized into three main layers: LLM infrastructure, foundation model, and downstream ecosystem.

\noindent \textbf{LLM Infrastructure~(\autoref{sec:infrastructure}).} The LLM infrastructure layer encompasses three critical areas: computational resources~(\autoref{sec:computational}), datasets~(\autoref{sec:dataset}), and development toolchain~(\autoref{sec:toolchain}). The computational resources area involves investigating hardware vulnerabilities and potential attacks, analyzing risks associated with distributed training systems, and developing strategies for efficient resource management. The development toolchain examination requires a deep dive into potential vulnerabilities and their propagation throughout the system, including the study of open-source software registries and repositories. The dataset aspect focuses on cleaning and curation methodologies, strategies to prevent data poisoning, and the complex landscape of data licensing and management.

\noindent \textbf{Foundation Model~(\autoref{sec:model}).} The foundation model layer addresses the core aspects of LLM development, testing, release, and deployment. This layer begins with in-depth research into model alignment during development \& training~(\autoref{sec:training}). It then progresses to comprehensive testing \& evaluation~(\autoref{sec:testing}), encompassing performance evaluation, reliability assessment, and safety testing. Following this, the release \& sharing phase~(\autoref{sec:release}) emphasizes model dependency analysis, risk propagation studies, and license compatibility investigations. Finally, the deployment \& maintenance phase~(\autoref{sec:deploy}) addresses the challenges of model drift and strategies for maintaining LLM performance in production environments.

\noindent \textbf{Application Ecosystem~(\autoref{sec:ecosystem}).} The downstream ecosystem explores the applications and systems built upon foundation models, with a focus on chatbot applications~(\autoref{sec:chatbot}) and agent-based systems~(\autoref{sec:agent}). The chatbot research area aims to uncover app security vulnerabilities and address governance issues in LLM-powered app stores. The agent-based systems area tackles challenges such as agent autonomy escalation, vulnerable agent logic, and various attack vectors.

\section{LLM Infrastructure}
\label{sec:infrastructure}
The model infrastructure is a foundational component of the LLM supply chain, encompassing the computational resources, datasets, and toolchains necessary for the training, testing, deployment, and maintenance of LLMs.

\subsection{Infrastructure: Computational Resources}
\label{sec:computational}
At the heart of model infrastructure are the computational resources, which include hardware such as graphics processing units (GPUs), tensor processing units (TPUs), and other specialized AI accelerators. These hardware components are complemented by software frameworks, distributed training frameworks, and cloud computing platforms that enable efficient utilization and management of these resources.
Our vision, encompassing both SE and S\&P perspectives, aims to establish a computational infrastructure that is resilient, scalable, and inherently secure. 
% By integrating cutting-edge SE methodologies with advanced security protocols, this approach ensures that the computational resources not only power the development of sophisticated models but also fortify the integrity of the entire LLM supply chain.

% \noindent \textbf{Vision [\underline{SE Perspective}].} 
From an SE perspective, our vision is the seamless integration of robust SE practices into the management and optimization of computational resources for LLM development. It emphasizes the adoption of \textit{\underline{Infrastructure as Code~(IaC)}} principles to enable version-controlled, reproducible, and scalable infrastructure setups. Emphasis is placed on creating highly scalable and flexible architectures that can efficiently handle the ever-increasing demands of LLM training and inference. This includes the development of intelligent resource allocation systems and orchestration tools that dynamically adjust to workload variations, ensuring optimal utilization of GPUs, TPUs, and other specialized AI accelerators. The vision also encompasses the implementation of advanced monitoring and diagnostics tools, enabling real-time performance tracking and predictive maintenance to minimize downtime and maximize resource efficiency.

% \noindent \textbf{Vision [\underline{S\&P Perspective}].} 
From an S\&P perspective, our vision champions a comprehensive ``\textit{\underline{defense-in-depth}}'' approach to securing the computational resources in the LLM supply chain. This holistic strategy addresses vulnerabilities at every level, from hardware to cloud infrastructure and software frameworks. At the hardware level, we envision the development of robust security measures to mitigate risks associated with over-reliance on specific vendors and hardware vulnerabilities. For cloud infrastructure, our vision emphasizes the development of advanced security frameworks specifically tailored for AI workflows. This includes fine-grained access control mechanisms and secure multi-tenant architectures. In the realm of software frameworks, we envision the integration of \textit{\underline{security-by-design}} principles into distributed training systems. This includes rigorous input validation, secure serialization protocols, and sandboxed execution environments for user-defined functions.

\noindent \textbf{Current State.}
The computational resources landscape in the LLM supply chain is characterized by a complex interplay of hardware advancements, cloud infrastructure, and software frameworks. NVIDIA's GPUs dominate the market~\cite{nvidia2024dominance}, leading to potential supply chain vulnerabilities due to over-reliance on a single vendor~\cite{eclypsium2024hardwareattacks,ionut2021gpumalware,trailofbits2024LeftoverLocals}. As an example of hardware vulnerabilities in the LLM supply chain, NVIDIA addressed a critical vulnerability (CVE-2023-31029) in NVIDIA DGX A100 baseboard management controllers (BMCs)~\cite{nvidia2024bulletin}, which allows unauthenticated attackers to execute arbitrary code and tamper with data remotely, potentially enabling the installation of persistent backdoors in the BMC firmware~\cite{eclypsium2024hardwareattacks}. Such backdoors could control all software executed on the server, including LLM models, training datasets, and model outputs, highlighting the severe implications of hardware vulnerabilities in the LLM supply chain.
The trend towards larger and more complex models has led to the widespread adoption of distributed computing systems and specialized AI cloud services. Major cloud providers like AWS, Google Cloud, and Azure have become integral to the LLM infrastructure, offering specialized AI services that, while enabling scalability, introduce new security risks~\cite{jair2024cloudrisks,shaerul2024cloudrisks}. On the software front, distributed training frameworks such as Horovod and PyTorch Distributed have enabled efficient large-scale model training but have also introduced new attack vectors. Recently, a critical vulnerability~(CVE-2024-5480) has been discovered in PyTorch's distributed RPC system. This vulnerability, stemming from insufficient input validation, could allow remote code execution when worker nodes serialize and send PythonUDFs (User Defined Functions) to another node without proper restrictions. This flaw potentially enables attackers to execute arbitrary code on affected systems, leading to unauthorized access to sensitive data, manipulation of model training processes, and compromise of the entire distributed computing environment~\cite{valdez2024rpc}. 

\noindent \textbf{Challenge I [\underline{S\&P Perspective}]: Hardware Vulnerabilities and Attacks.}
Reliance on proprietary hardware solutions introduces vendor lock-in risks and supply chain vulnerabilities. Ensuring compatibility, optimizations, and security across the hardware stack is challenging, particularly when integrating components from diverse suppliers. The proprietary nature of many hardware designs and firmware creates a ``black box'' scenario, where potential vulnerabilities or backdoors~\cite{zhu2017gpubackdoor,yang2016hardwarebackdoor,gohil2022hardwaretrojan} may remain undetected. Moreover, side-channel attacks pose a significant threat, particularly in the context of GPUs. Due to their parallel processing nature and shared resources, GPUs can leak sensitive information, including model parameters~\cite{naghibijouybari2018gpuperformancecounters}, through timing differences~\cite{jiang2019gputiming}, power consumption patterns~\cite{taneja2023hot}, or electromagnetic emissions~\cite{zhan2022gpuem}. The complexity of GPU architectures and their proprietary nature further complicate the implementation of comprehensive security measures against such attacks.
These challenges provide some opportunities and future research directions:

\begin{itemize}[leftmargin=15pt]
    \item \textbf{Opportunity: Secure Hardware Design.} The unique requirements of LLM training and inference present an opportunity to rethink hardware design with security as a fundamental principle. This could involve incorporating security features at the hardware level~\cite{zhu2020gputee,mai2023gputee}, designed to resist side-channel attacks and other hardware-level threats. Such secure-by-design hardware could provide a strong foundation for building trustworthy AI, ensuring the integrity and confidentiality of LLM operations from the hardware level up.
    
    \item \textbf{Opportunity: Hardware Trojan/Backdoor Mitigation.} The complexity of hardware-level threats opens up opportunities for developing sophisticated detection and mitigation mechanisms. Current approaches primarily fall into two categories~\cite{gohil2022hardwaretrojan}: logic testing and side-channel analysis. Logic testing involves applying input patterns to the chip and monitoring output responses to detect deviations from expected behavior, crucially relying on Trojan-free~(or golden) chips as a reference~\cite{cahkraborty2009golden,pan2021rlgolden,vashistha2022nogolden}. Side-channel analysis, in contrast, focuses on monitoring physical characteristics such as power consumption~\cite{zhao2020power,salmani2012power}, path delay~\cite{yier2008pathdelay}, or electromagnetic emissions~\cite{agrawal2007fingerprint,he2017nogoldenem} to identify anomalies that might indicate the presence of a Trojan. Both methods face challenges such as requiring numerous input patterns~\cite{valdez2024rpc,pan2021rlgolden}, low detection rates~\cite{valdez2024rpc,cahkraborty2009golden}, and, for logic testing, reliance on hard-to-obtain golden chips~\cite{he2017nogoldenem,vashistha2022nogolden}. To address these challenges, future research could focus on developing more efficient and accurate detection methods that reduce dependence on golden chips and minimize the number of required input patterns. Advanced AI techniques, particularly reinforcement learning~\cite{gohil2022rltrojan} and LLM~\cite{chaudhuri2024spiced,kokolakis2024harnessing}, show potential for improving detection accuracy and efficiency. Future research in these areas may lead to improvements in hardware security practices, contributing to the integrity and trustworthiness of LLM systems from the hardware level up.
\end{itemize}

\noindent \textbf{Challenge II [\underline{S\&P Perspective}]: Distributed Training Risks.} 
The advent of LLMs has necessitated the adoption of distributed training approaches, introducing a complex array of security challenges and supply chain risks. This distributed infrastructure, while enabling unprecedented model capabilities, significantly expands the attack surface, with each node representing a potential vulnerability~\cite{lyu2020threats}. Node failures, whether due to hardware malfunctions, network issues, or targeted attacks, pose critical risks to training continuity and data integrity~\cite{liu2023distributedfailure,verbraeken2020distributedsurvey}. Moreover, the complexity of distributed training frameworks, such as PyTorch Distributed, Horovod, and TensorFlow Distributed, also expand the attack surface significantly. First, there is the potential for inadequate encryption or authentication mechanisms, which could expose critical components like gradient updates and model parameters to interception or tampering by malicious actors. Second, the complexity of these systems can sometimes lead to misconfigurations, such as exposed APIs or insufficient access controls~\cite{valdez2024rpc,rayattacks}, which may create additional security vulnerabilities.

\begin{itemize}[leftmargin=15pt]
    \item \textbf{Opportunity: Systematic Vulnerability Analysis of Distributed Training Frameworks.} Despite the widespread adoption of frameworks such as PyTorch Distributed, Horovod, and TensorFlow Distributed for large-scale model training, there is a notable absence of comprehensive studies that systematically catalog and analyze their security risks, threat landscapes, and potential attack vectors. As distributed training operations grow in scale and complexity, the full spectrum of vulnerabilities in their frameworks remains largely uncharted. This vast and poorly understood threat landscape encompasses unexplored attack surfaces stemming from the frameworks' unique operational characteristics, including high-volume inter-node communications, complex synchronization mechanisms, and potential cascading failures across the distributed system. This research gap not only hampers the development of effective security measures but also leaves practitioners without clear guidelines for securing their distributed training environments. 
\end{itemize}

\noindent \textbf{Challenge III [\underline{SE Perspective}]: Computational Resource Management and Optimization.}
The computational resources in the LLM supply chain present unique challenges and opportunities from an SE perspective. One significant challenge is the need for efficient resource allocation and optimization in distributed training environments. As LLMs grow in size and complexity, traditional SE practices struggle to manage the massive parallelism and data flow required for training. This necessitates the development of novel scheduling algorithms and load-balancing techniques that can dynamically adapt to the varying computational demands of different training phases. Another challenge lies in ensuring the reproducibility and version control of training environments, as subtle differences in hardware configurations or software dependencies can lead to inconsistent model behaviors.

\begin{itemize}[leftmargin=15pt]
    \item \textbf{Opportunity: Advancing IaC and Intelligent Compute Scheduling.} The challenges in LLM resource management present an opportunity to advance IaC practices and develop intelligent compute scheduling systems tailored for AI workflows~\cite{diaz2024iac}. This involves creating sophisticated IaC tools that can dynamically provision and manage the complex, heterogeneous infrastructure required for LLM training. Such tools would enable version-controlled, reproducible environment setups across different cloud providers and on-premises systems, ensuring consistency in training results.
    
    \item \textbf{Opportunity: Green SE Practices for LLM Training.} The increasing focus on the environmental impact of AI training presents an opportunity to implement green SE practices specifically for LLM development~\cite{shi2024green}. This involves creating methodologies and tools to optimize energy consumption without compromising model performance. Potential areas of innovation include the development of energy-aware scheduling algorithms that prioritize workloads based on both performance requirements and energy efficiency. Software engineers could focus on creating more efficient model architectures and training algorithms that reduce computational requirements while maintaining model quality.
\end{itemize}

\subsection{Infrastructure: Datasets}
\label{sec:dataset}
The datasets employed for training play a crucial role in the development of LLMs~\cite{jain2020overview,Nithya2021datacascades}, which encompass vast collections of text from various sources, including books, websites, academic papers, social media, and open-source repositories. These datasets are complemented by data processing pipelines, annotation tools, and data management systems that enable efficient curation, cleaning, and utilization of the data.
Our vision aims to establish a data infrastructure that is diverse, scalable, inherently secure, and privacy-preserving~\cite{mazumder2024dataperf}.

From an SE perspective, this vision aims to improve dataset management in the LLM supply chain through advanced SE practices. It emphasizes creating highly efficient and flexible data processing pipelines that can handle the ever-increasing volume and diversity of training data. This includes the development of intelligent data curation systems and annotation tools that dynamically adapt to evolving data quality requirements, ensuring optimal utilization of human and computational resources in data preparation. The vision also encompasses the implementation of advanced data provenance and lineage tracking tools, enabling comprehensive auditing and reproducibility of dataset creation and modification processes.

From an S\&P perspective, our vision for dataset management in the LLM supply chain embraces an \textit{\underline{inherent security and privacy}} approach, embodying the principle of \textit{\underline{shifting left}}~\cite{codacy_shift_left_security} these critical considerations. This paradigm aims to address potential risks at the earliest stages of data collection and curation, rather than relying solely on downstream mitigation strategies. We envision the development of advanced content filtering and toxicity detection systems capable of identifying and removing harmful, biased, or privacy-infringing content during the initial stages of dataset creation.  Furthermore, we aim to implement privacy-preserving data collection methods that respect individual consent and data rights from the outset, ensuring compliance with privacy regulations such as GDPR and CCPA at the point of data ingestion. 
% By embedding these security and privacy considerations directly into the fabric of dataset creation and management, we aim to build a foundation for LLM development that is intrinsically more secure, ethical, and respectful of individual privacy.

\noindent \textbf{Current State.}
The current landscape of datasets in the LLM supply chain is characterized by unprecedented scale and diversity, coupled with growing concerns about data quality, bias, and privacy. Recent studies have shown that LLMs can perpetuate and amplify societal biases present in the source data~\cite{kirk2021bias,yu2024large,navigli2023biases}. Additionally, the inclusion of copyrighted material~\cite{meeus2024copyright,li2024digger} and privacy information~\cite{carlini2021extracting,carlini2023extracting} in training datasets has also become a contentious legal issue. In the realm of Code LLMs, publicly available code repositories have become crucial sources of training data, presenting unique challenges related to code quality~\cite{jain2023llm}, licensing~\cite{majdinasab2024trained,katzy2024codellmlicense}, and potential security vulnerabilities~\cite{yang2024memory} in the training data itself. What's more, security vulnerabilities in data processing pipelines have emerged as a critical concern~\cite{zhao2024models}, highlighting the need for more rigorous security practices in dataset preparation and model training workflows. For instance, the discovery of CVE-2024-39705 in NLTK~\cite{smartkeyss2024nltkrce} and NVDB-CNVDB-2023879241 in Hugging Face~\cite{tencent2024datasets} revealed potential arbitrary code execution vulnerabilities, posing significant risks in dataset processing. As the field continues to evolve rapidly, addressing these challenges in dataset management, security, and privacy remains crucial for the responsible development of future LLM technologies, both for natural language and code-related applications.

% ~\shenao{Need Another Challenge About This!}

% \begin{table}[htbp]
% \centering
% \small
% \begin{tabular}{|p{0.1\textwidth}|p{0.27\textwidth}|p{0.27\textwidth}|p{0.27\textwidth}|}
% \hline
% & \textbf{Vision} & \textbf{Challenges} & \textbf{Opportunities} \\
% \hline
% \textbf{SE Perspective} & 
% \begin{itemize}[leftmargin=*,nosep]
% \item Efficient dataset pipelines
% \item Intelligent data curation
% \item Advanced data provenance tracking
% \end{itemize} &
% \begin{itemize}[leftmargin=*,nosep]
% \item Data cleaning and curation
% \item Handling diverse data sources
% \item License management complexity
% \end{itemize} &
% \begin{itemize}[leftmargin=*,nosep]
% \item Automated license conflict auditing
% \item Complex license understanding tools
% \end{itemize} \\
% \hline
% \textbf{S\&P Perspective} & 
% \begin{itemize}[leftmargin=*,nosep]
% \item Inherent security and privacy in the collected dataset
% \end{itemize} &
% \begin{itemize}[leftmargin=*,nosep]
% \item Handling privacy and PII
% \item Mitigating bias and toxicity
% \item Avoiding data poisoning
% \end{itemize} &
% \begin{itemize}[leftmargin=*,nosep]
% \item Privacy-preserving algorithms
% \item Bias detection and mitigation
% \item Robust data validation techniques
% \end{itemize} \\
% \hline
% \end{tabular}
% \caption{Vision, Challenges, and Opportunities in Datasets}
% \label{tab:datasets}
% \end{table}

\noindent \textbf{Challenge I~[\underline{SE + S\&P Perspective}]: Dataset Cleaning and Curation.}
The process of data cleaning and curation is a critical step in the development of LLMs, serving as the backbone for ensuring integrity, privacy, and ethical alignment. 
The primary obstacles stem from the handling of privacy~\cite{carlini2021extracting,kim2024propile,lukas2023analyzing}, biased~\cite{kirk2021bias,yu2024large,ferrara2023should,zhuo2023red}, and toxic~\cite{weidinger2022taxonomy,zhuo2023red,ousidhoum2021probing} data in training sets, each of which presents unique challenge.
The potential privacy challenges are twofold: ensuring that personally identifiable information~(PII) is not present in the training data~\cite{brown2022does} and preventing the model from learning to reproduce or infer it from the patterns it is trained on~\cite{carlini2021extracting,carlini2023extracting,kim2024propile}. 
Bias in training data is a well-documented issue that can lead models to perpetuate or even amplify existing prejudices~\cite{steed2022upstream,bommasani2021opportunities,liang2021towards}. The challenge lies in identifying and mitigating biases, which are often deeply embedded in the data and reflective of broader societal biases~\cite{gallegos2023bias,steed2022upstream}. The presence of toxic and harmful content in training datasets poses a significant risk to the safety and reliability of LLMs~\cite{weidinger2022taxonomy,zhuo2023red,ousidhoum2021probing,birhane2024into}. Models trained on datasets containing such content may reproduce or even generate harmful outputs, undermining their applicability in diverse contexts. These challenges in data cleaning and curation require sophisticated strategies for mitigation, and this provides some opportunities as discussed below.

\begin{itemize}[leftmargin=15pt]
    % \item \textbf{Opportunity: Deduplication.} At the forefront of this opportunity is the development of more sophisticated deduplication algorithms. Simple deduplication methods such as MinHash~\cite{broder1997resemblance} often struggle with the scale and diversity of data typical for LLM training~\cite{cheng2021lofs}. Advanced deduplication strategies that carefully evaluate which duplicates to remove can ensure that the richness of the data is maintained. There lies a potential in leveraging careful data selection via pre-trained model embeddings, ensuring that training data are both diverse and concise. Innovations in this area could significantly reduce computational overhead and improve model performance.

    \item \textbf{Opportunity: Privacy Preserving.} In the realm of privacy preservation for training LLMs, several techniques (e.g., text anonymization~\cite{pilan2022text} and differential privacy techniques~\cite{hassan2019differential}) have emerged. However, challenges persist, particularly regarding knowledge integrity; removing sensitive information may lead to the loss of critical context, potentially misleading LLMs. Research focused on balancing anonymization and utility is essential. Additionally, the complexity of multimodal data complicates effective anonymization across diverse data types, highlighting the need for dedicated multimodal privacy frameworks.

    % The development and implementation of innovative privacy preserving algorithms stand out as a primary opportunity. Current methods such as xxx have set the foundation, yet they often face challenges in balancing privacy with data utility. The need to preserve privacy while ensuring that the dataset remains comprehensive and informative enough to train robust models is still an open problem.
    
    % Privacy-preserving techniques such as differential privacy offer a pathway forward~\cite{shi2022just,shi2021selective,plant2022you,duan2024flocks}, but their integration into the data curation process requires careful balancing among fairness and transparency~\cite{rust2023differential,srivastava2024amplifying};
    
    \item \textbf{Opportunity: Bias Mitigation.} The first opportunity lies in enhancing methodologies for the detection and correction of biases in datasets. While significant progress has been made~\cite{park2018reducing,navigli2023biases,liu2022quantifying,wang2024exploring}, there is a continuous need for more sophisticated tools that can identify subtle and complex biases. Another critical opportunity is to strike a balance between removing biases and maintaining the representativeness of datasets. This involves not only the removal of harmful biases but also ensuring that the diversity and richness of human experiences are accurately reflected in LLMs. 
    % Techniques such as data augmentation and synthetic data generation can help achieve this balance, ensuring that LLMs are both fair and comprehensive in their understanding of the world.
    % Efforts to address this involve developing methodologies for bias detection and correction~\cite{park2018reducing,navigli2023biases,liu2022quantifying}. However, implementing these methodologies without compromising the dataset's representativeness or introducing new forms of bias is an ongoing challenge~\cite{gallegos2023bias,ferrara2023should,zhou2020challenges,steed2022upstream,raza2024fair}.
    
    \item \textbf{Opportunity: Detoxifying.} Cleaning datasets of toxic content such as hate speech, misinformation, and abusive language requires not only sophisticated detection tools~\cite{kopf2024openassistant,zhang2024efficient,bespalov2023towards} but also a nuanced understanding of the mechanism what constitutes harm~\cite{welbl2021challenges,pavlopoulos2020toxicity,baldini2021your}, which can vary widely across different cultural and social contexts. Cross-cultural sensitivity presents an opportunity to create guidelines and frameworks that respect cultural differences while identifying universally harmful content. 
    
    % Developing culturally sensitive models requires a concerted effort to diversify the datasets used for training and to involve a broad spectrum of voices in the model development process.
\end{itemize}

\noindent \textbf{Challenge II~[\underline{S\&P + SE Perspective}]: Avoid Dataset Poisoning.} 
Data poisoning attacks~\cite{zhai2023text,schwarzschild2021just,OWASP-LLM} pose severe supply chain risks for LLMs, as attackers can degrade model performance or introduce backdoors~\cite{li2021hidden,yang2023data,abdali2024securing,li2023multi} through corrupted training data, which undermines the integrity and reliability of LLMs. Additionally, supply chain attacks targeting the data storage, processing, or distribution infrastructure can facilitate data poisoning or corruption, potentially compromising the entire model development lifecycle. Avoiding data poisoning in the supply chain of LLMs presents a multifaceted set of challenges, intricately linked with the broader objectives of data cleaning and curation.

\begin{itemize}[leftmargin=15pt]
    \item \textbf{Opportunity: Data Validation.} The first line of defense against data poisoning is data validation~\cite{polyzotis2019data,hutchinson2021towards}, a process that is inherently complex due to the vast scale and heterogeneity of datasets used in LLM training. Effective validation requires sophisticated algorithms capable of detecting anomalies and malicious modifications in the data~\cite{lwakatare2021experiences}, which is a task that becomes exponentially difficult as the data volume and diversity increase~\cite{paleyes2022challenges,shivashankar2022maintainability}. The opportunity for progress in robust data validation resides in advancing algorithmic solutions that are capable of nuanced detection of subtle and sophisticated data manipulation attempts.  These solutions must be scalable enough to manage the expansive datasets characteristic of LLM training, thereby ensuring comprehensive coverage without compromising efficiency.
    
    \item \textbf{Opportunity: Provenance Tracking.} Provenance tracking, or the ability to trace the origin and history of each data point, becomes paramount in a landscape where data can be compromised at any stage~\cite{xin2021production,rupprecht2020improving,tang2019sac,hutchinson2021towards}. Implementing such tracking mechanisms involves not only technical solutions~\cite{namaki2020vamsa} but also organizational policies that ensure data sources are reputable and that data handling practices are transparent~\cite{foundationmodeltransparencyindex} and secure. However, establishing a provenance tracking system that is both comprehensive and efficient remains an open problem, given the complexity of LLM supply chains and the potential for data to be aggregated from myriad sources~\cite{longpre2023data,zhang2020said,wang2015big}.
    
\end{itemize}

% \shenao{Review\#3: Challenge III: The description of the challenge should be more concrete. The text is abstract and unclear to me.}

\noindent \textbf{Challenge III~[\underline{SE Perspective}]: Dataset Licensing and Management.} Dataset licensing and management present several concrete challenges critical for maintaining legal and ethical standards in the development of LLMs. As these models require vast amounts of diverse data for training, the risk of copyright infringement and licensing violations becomes significant. For instance, using datasets without proper licenses can lead to legal liabilities, including lawsuits and financial penalties. Recent research has highlighted specific issues such as the ambiguity surrounding the licensing of web-scraped data and the complexities involved in determining dataset ownership~\cite{katzy2024exploratory,li2024digger,chu2024protect,karamolegkou2023copyright}. Furthermore, the diversity of data sources—from open-access repositories to proprietary databases—complicates compliance with varying licensing terms. The often opaque legal frameworks governing data use exacerbate these challenges, making it difficult for developers to navigate the landscape effectively~\cite{foundationmodeltransparencyindex}. 
% Addressing these challenges opens up opportunities for further research into developing best practices for dataset licensing management, creating tools for automated compliance checks, and establishing clearer guidelines for data usage in LLM training.

% \noindent \textbf{Challenge III~[\underline{SE Perspective}]: License Management.}
% License management encompasses a range of challenges that are critical to navigate in order to maintain legal and ethical standards. As LLMs require vast amounts of diverse data for training, the risk of copyright infringement, licensing violations, and subsequent legal liabilities intensifies~\cite{katzy2024exploratory,li2024digger,chu2024protect,karamolegkou2023copyright}. Recent research~\cite{wolter2023open,longpre2023data,wu2001open,vendome2015large,sun2022coprotector} has shed light on the complex landscape of dataset copyright and licensing, underscoring the need for further exploration and development of best practices. This need is further complicated by the diversity of data sources and the often opaque legal frameworks governing data use~\cite{foundationmodeltransparencyindex}. These challenges also open up some opportunities for further research.

\begin{itemize}[leftmargin=15pt]

    \item \textbf{Opportunity: Understanding Complex Licenses.} One of the primary challenges in dataset licensing management is the complexity and diversity of licenses~\cite{vendome2016assisting,vendome2015license,vendome2017license,di2010exploratory,vendome2015large}. Data sources can range from openly accessible datasets with permissive licenses to proprietary datasets with stringent usage restrictions. Each source comes with its own set of legal terms, necessitating careful review and comprehension to ensure compliance~\cite{wolter2023open,wu2017analysis,barcomb2022open,german2010sentence,vendome2017machine}. Opportunities in this area include developing automated tools for detecting and summarizing key legal terms, providing stakeholders with clear and accessible insights into the permissions, obligations, and restrictions associated with each dataset.

    \item \textbf{Opportunity: Automated License Conflict Auditing.} Implementing automated license conflict auditing systems represents a significant opportunity to enhance dataset licensing management practices~\cite{liu2024datasets,albalak2024survey,tuunanen2009automated}. Such systems could streamline the verification process for compliance with licensing agreements across extensive datasets. However, developing these systems presents technical challenges, including the need for advanced algorithms capable of interpreting and applying the legal nuances of various licenses~\cite{van2014tracing}. 

\end{itemize}

\subsection{Infrastructure: LLM Development Toolchain}
\label{sec:toolchain}
In the realm of LLMs, the development tools and frameworks serve as the cornerstone of innovation, significantly shaping the trajectory of artificial intelligence. This vision, from both SE and S\&P standpoints, is ambitious yet grounded, aiming to forge a development environment that is robust, scalable, and inherently secure. By weaving together the best practices of SE with advanced security measures, this approach ensures that the toolchain not only enable the crafting of sophisticated models but also safeguard the integrity of the entire supply chain.

% \noindent \textbf{Vision [\underline{SE Perspective}].} 
From an SE perspective, central to this vision is the seamless incorporation of SE best practices into LLM development tools and frameworks. Modular design principles are prioritized to boost maintainability and scalability, allowing for seamless updates and modifications without impacting the broader system. The vision also encompasses the implementation of continuous integration and deployment (CI/CD) pipelines to streamline testing and deployment processes, along with comprehensive tooling that empowers developers to efficiently implement and leverage these best practices in their workflows.

% \noindent \textbf{Vision [\underline{S\&P Perspective}].} 
From an S\&P perspective, a ``\textit{\underline{security by design}}'' philosophy is advocated, embedding security considerations at the onset of the development process. This includes deploying comprehensive code analysis tools for early vulnerability detection and enforcing secure authentication. Beyond individual security practices, a crucial aspect of safeguarding the LLM supply chain involves addressing the security of the development tools' own supply chains. Given the reliance of most LLM systems on a handful of core frameworks, the compromise of any one of these foundational elements could expose a vast array of LLM systems to risk~\cite{rayattacks,liu2024datasets,safetensortransvul}. To mitigate these risks, the vision calls for rigorous security measures at every level of the supply chain for development tools and frameworks. 

\noindent \textbf{Current State.} The development toolchain for LLMs is characterized by a rapidly evolving ecosystem of third-party libraries, frameworks, and specialized tools designed to address the unique challenges of LLM development. As discussed in \autoref{sec:computational} and \autoref{sec:dataset}, distributed training frameworks and data processing pipelines introduce a set of security concerns. Beyond these, Hugging Face's Transformers library has emerged as a de facto standard, providing a comprehensive suite of pre-trained models and fine-tuning tools. However, its widespread adoption has raised concerns about over-reliance on a single framework, potentially introducing systemic vulnerabilities across the LLM supply chain~\cite{tencent2024datasets,zhao2024models}. Major deep learning frameworks like PyTorch and TensorFlow continue to play crucial roles, but they are not immune to security issues. For instance, TensorFlow's Keras framework suffered from a Lambda Layer vulnerability~(CVE-2024-3660) allowing arbitrary code injection~\cite{cert2024keras,keras2024downgrade,hiddenlayer2023keras}, while PyTorch's use of Pickle for model serialization introduces potential deserialization vulnerabilities~\cite{zhao2024models,trailofbits2024pickle1,trailofbits2024pickle2}. The toolchain landscape has also seen the emergence of specialized tools addressing specific aspects of LLM development, such as Weights \& Biases~\cite{weight2024platform} for experiment tracking and Ray~\cite{ray2024platform} for distributed computing. These tools, while enhancing productivity, introduce new potential attack vectors~\cite{rayattacks,blackhat2024mlops,blackhat2024practical}. A recent path traversal vulnerability (CVE-2023-1177) in MLflow~\cite{mlflow2024cve}, a popular experiment tracking tool, allowed unauthorized access to files outside the intended directory. This flaw risked exposing sensitive data and configuration files, underscoring the importance of proper input validation in LLM infrastructure tools. The increasing complexity of the toolchain, coupled with the rapid pace of development, has created challenges in maintaining consistent security standards across the LLM development pipeline.

\noindent \textbf{Challenge I~[\underline{SE Perspective}]: Understand LLM Development Toolchain.}
The rapid evolution and increasing complexity of the LLM development toolchain pose significant challenges for developers and security researchers alike. Understanding the intricate interplay between various components, from pre-processing libraries to distributed training frameworks, requires a deep dive into their functionalities. This challenge is further compounded by the fast-paced nature of LLM research, where new tools and techniques are constantly emerging, making it difficult to maintain a comprehensive and up-to-date understanding of the entire toolchain ecosystem.

\begin{itemize}[leftmargin=15pt]
    \item \textbf{Opportunity: LLM Development Toolchain Mining.} A promising avenue for enhancing supply chain security lies in the opportunity of LLM development toolchain mining, which involves the systematic analysis and evaluation~\cite{jiang2022empirical} of the tools and libraries used in the lifecycle of LLMs. The core of this opportunity revolves around the comprehensive mining and auditing of development tools, from code libraries to data processing frameworks used in LLM training. Through the detailed analysis of the toolchain, developers can identify redundancies, inefficiencies, and areas for improvement, paving the way for the development of more streamlined, effective, and secure LLMs. Additionally, this mining process can spur innovation by highlighting gaps or needs within the toolchain, driving the creation of new tools, or the enhancement of existing ones to better serve the evolving demands of LLM development.
\end{itemize}

\noindent \textbf{Challenge II~[\underline{S\&P Perspective}]: Potential Toolchain Vulnerability.}
As the LLM toolchain grows in complexity, conducting systematic vulnerability analyses becomes increasingly challenging~\cite{wu2024new,liu2023demystifying,he2024security}. Each component of the toolchain, from data preprocessing to model serving, introduces potential security risks. The challenge lies in developing comprehensive methodologies for identifying, assessing, and mitigating vulnerabilities across this diverse ecosystem. 
% The interconnected nature of these tools further complicates the analysis, as vulnerabilities in one component may have cascading effects throughout the toolchain.

\begin{itemize}[leftmargin=15pt]
    \item \textbf{Opportunity: Systematic Toolchain Vulnerability Analysis.} The complexity of the LLM development toolchain presents an opportunity for innovative approaches to security analysis. By developing systematic methods for vulnerability assessment across the entire toolchain, we can significantly enhance the overall security posture of LLM development processes. This opportunity encompasses the creation of comprehensive vulnerability scanning tools tailored specifically for LLM development environments. These tools would need to understand the unique characteristics of AI frameworks, distributed computing systems, and data processing pipelines used in LLM development. Additionally, the establishment of \underline{a centralized vulnerability database} specific to LLM development tools could serve as a valuable resource for developers and security researchers, facilitating rapid identification and mitigation of known vulnerabilities. Such systematic approaches to toolchain vulnerability analysis could lead to more robust and secure LLM development processes.
\end{itemize}

\noindent \textbf{Challenge III~[\underline{S\&P + SE Perspective}]: Intricate Dependency and Vulnerability Propagation.}
Managing the intricate dependency tree, including both open-source and commercial components, poses a significant challenge within the supply chain~\cite{liu2022demystifying,hejderup2018software,ma2018constructing}. Supply chain attacks targeting the development infrastructure or code repositories could lead to the injection of vulnerabilities or malicious code~\cite{li2023malwukong,duan2020towards,guo2023empirical}, potentially compromising the entire lifecycle of model development and deployment. Moreover, vulnerabilities within dependencies or components can propagate through the supply chain~\cite{hu2023empirical,zhang2023mitigating}, adversely affecting the security and reliability of models. Establishing robust dependency management processes, conducting thorough security monitoring, and ensuring supply chain transparency are essential for mitigating risks such as compromises in the LLM development tools supply chain.

\begin{itemize}[leftmargin=15pt]
    \item \textbf{Opportunity: Development of SBOM for LLM Toolchain.} The adoption of the software bill of materials~(SBOM) of LLM toolchain presents a unique opportunity to achieve unprecedented levels of transparency and security. By meticulously documenting every library, dependency, and third-party component, SBOM enables developers to gain a comprehensive overview of their tools' software ecosystem~\cite{xia2023empirical,stalnaker2024boms}. This holistic visibility is instrumental in identifying vulnerabilities, outdated components, and non-compliant software elements that could jeopardize the development process and, ultimately, the security of the LLMs themselves. The detailed insights provided by SBOMs pave the way for proactive vulnerability management. Armed with knowledge about constituent components, development teams can swiftly address security flaws, apply necessary patches, and update components. This preemptive identification and remediation process is crucial in safeguarding models against potential exploits that could jeopardize their reliability and the security of the systems they operate within.
    
    \item \textbf{Opportunity: Dependency Mapping and Impact Analysis.} There is an opportunity for researchers to focus on developing advanced methodologies for dependency mapping and impact analysis within the LLM development toolchain. By creating comprehensive models that visualize and analyze the intricate relationships between various dependencies, researchers can better understand how vulnerabilities propagate through the supply chain. This research can lead to the formulation of best practices for dependency management, enabling teams to assess the risk associated with each component more effectively. Additionally, such insights can inform strategies for mitigating the effects of compromised dependencies, thereby enhancing the overall security and reliability of LLM systems.
\end{itemize}

\section{Foundation Model}
\label{sec:model}
In the evolving landscape of LLMs, the vision for the foundation model within the supply chain encompasses a holistic and agile approach, from initial development to deployment, maintenance, and updates. This lifecycle is envisioned to be a seamless continuum that not only addresses the inherent challenges but also leverages them as catalysts for innovation and progression in the field.

\subsection{Model: Development \& Training}
\label{sec:training}
The vision for developing and training LLMs is a compelling narrative of innovation, inclusivity, and ethical responsibility, aiming to push the boundaries of what these computational behemoths can achieve while grounding their evolution in principles that benefit all of humanity. This vision integrates the cutting edge of technological innovation with an unwavering commitment to ethical principles and operational efficiency. 

From an SE perspective, the vision for LLM development and training is centered on achieving unprecedented efficiency and environmental sustainability. As the computational demands of these models continue to soar, innovative approaches to training are prioritized. This future envisions the development of highly efficient algorithms that dramatically reduce the computational resources required for training, making the development of state-of-the-art models more accessible and environmentally friendly. 

From an S\&P perspective, the vision advocates for a comprehensive approach to security and ethical responsibility, integrated from the very outset of model development and training. This approach ensures that LLMs are developed with a deep understanding of their potential societal impacts, embedding robust safeguards against all forms of harmful content into the core of model architecture and training processes, including but not limited to misinformation, hate speech, explicit content, and malicious instructions. This holistic approach to security and ethics aims to create LLMs that are not only powerful but also trustworthy, safe, and beneficial to society.

\noindent \textbf{Current State.} The current state of fundamental model development and training in the LLM supply chain, while advanced, still faces significant challenges in realizing the full vision. While ethical considerations and bias mitigation strategies are increasingly recognized as crucial, their implementation frequently occurs as a post-hoc process rather than being intrinsically woven into the development fabric. Consequently, many current models exhibit biases and alignment issues that necessitate ongoing monitoring and correction. The interpretability and transparency of fundamental models remain limited, with many operating as ``black boxes'', hindering effective debugging and alignment efforts~\cite{singh2024rethinking}. However, promising advancements are being made in this area~\cite{li2024glitch,zhang2024glitchprober,singh2024rethinking,tan2024sparsity}, bridging the gap between the current state and the envisioned future of fundamental model development and training.
 
\noindent \textbf{Challenge~[\underline{S\&P + SE Perspective}]: Model Alignment.}
As the capabilities of LLMs continue to expand, ensuring their alignment~\cite{liu2023trustworthy,shen2023large,wolf2023fundamental} with human values and intentions become increasingly critical. The concept of inner alignment~\cite{hubinger2019risks,shen2023large} focuses on ensuring that an LLM's objectives are congruent with the intentions of its designers during the development and training phase. Outer alignment~\cite{shen2023large}—ensuring that the model's outputs and behaviors align with societal norms and expectations—also plays a vital role in the overall effectiveness of LLMs.
The pursuit of both inner and outer alignment requires a multifaceted strategy. Inner alignment is complicated by nuanced failure modes~\cite{shen2023large,angelou2022three,hubinger2019risks}, such as proxy, approximate, and suboptimality alignments, each presenting unique challenges in ensuring LLM systems operate as intended. These failure modes underscore the potential divergence between a model's optimized objectives and the overarching goals its designers aim to achieve.
Simultaneously, outer alignment challenges arise from the diverse values and expectations present in society, necessitating ongoing engagement with stakeholders to define acceptable behaviors and outputs. Methodological approaches such as relaxed adversarial training~\cite{hubinger2019risks} and partitioning gradients~\cite{yu2023unlearning} have been proposed to enhance inner alignment. However, achieving effective outer alignment requires transparency in the LLM system's decision-making processes and continuous feedback from users and society at large.

% As the capabilities of LLMs continue to expand, the necessity of ensuring their alignment~\cite{liu2023trustworthy,shen2023large,wolf2023fundamental} with human values and intentions becomes increasingly critical. The concept of inner alignment~\cite{hubinger2019risks,shen2023large} focuses on ensuring that an LLM's objectives are congruent with the intentions of its designers during the development and training phase. The pursuit of inner alignment during the development and training phase of LLMs requires a multifaceted strategy. However, inner alignment is complicated by its nuanced failure modes~\cite{shen2023large,angelou2022three,hubinger2019risks}, such as proxy, approximate, and suboptimality alignments, each presenting unique challenges in ensuring LLM systems operate as intended. These failure modes underscore the potential divergence between a model's optimized objectives and the overarching goals its designers aim to achieve. 
% To address these issues, methodological approaches such as relaxed adversarial training~\cite{hubinger2019risks} and partitioning gradients~\cite{yu2023unlearning} have been proposed. However, the efficacy of such methodologies hinges on the transparency of the LLM system's decision-making processes, which provides opportunities for further research.

\begin{itemize}[leftmargin=15pt]
   \item \textbf{Opportunity: Advancing Interpretability of LLMs.} The opportunity to advance the methodology of transparency and interpretability in LLMs stands as a critical endeavor~\cite{singh2024rethinking}. Enhancing transparency involves shedding light on the often opaque decision-making processes of these models, enabling a clearer understanding of how inputs are processed and interpreted to produce outputs~\cite{zou2023representation}. By demystifying the inner workings of LLMs, researchers, and practitioners can gain valuable insights into the operational dynamics of these models, identifying areas where the models' behaviors may not align with expected or desired outcomes~\cite{singh2024rethinking,jiang2021alignment,he-etal-2024-llm}. This early detection is invaluable, as it allows for timely interventions to correct course~\cite{li2024inference,tan2024sparsity}, preventing minor misalignments from escalating into more significant issues.

    \item \textbf{Opportunity: Enhancing Feedback Mechanisms.} The integration of robust feedback mechanisms into LLMs represents a transformative opportunity to enhance their adaptability and alignment with human values over time~\cite{peng2023check,madaan2024self}. By embedding iterative feedback loops within the architecture of LLMs, developers can establish a dynamic process where the models continually learn and adjust from real-world interactions and user feedback. Feedback loops can be particularly beneficial in identifying and correcting biases, misconceptions, or inaccuracies that may emerge in LLM outputs, thereby enhancing the models' trustworthiness and reliability~\cite{ouyang2022training,peng2023check,liu2023chain}. This process enables LLMs to evolve and adapt in response to changing contexts, user needs, and societal norms, ensuring their ongoing relevance and utility.
\end{itemize}

\subsection{Model: Testing \& Evaluation}
\label{sec:testing}
The vision for testing and evaluating LLMs is a holistic approach that ensures these powerful models are not only technologically advanced but also ethically sound, socially beneficial, and robustly secure. This vision integrates rigorous technical evaluation with comprehensive ethical and societal impact assessments.

From an SE perspective, the vision for LLM testing and evaluation focuses on developing comprehensive, adaptive, and efficient assessment frameworks. This future envisions automated testing pipelines that can continuously evaluate models across a wide range of metrics, including performance, robustness, and generalization capabilities. The goal is to create dynamic evaluation systems that can evolve alongside rapid advancements in LLM capabilities, ensuring that testing methodologies remain relevant and effective.

From an S\&P perspective, the vision emphasizes a proactive and integrated approach to security, privacy, and ethical evaluation. This approach aims to embed ethical considerations and security assessments throughout the entire evaluation process, rather than treating them as separate concerns. The goal is to develop comprehensive frameworks that can rigorously test for biases, fairness issues, potential misuse scenarios, and vulnerabilities to attacks, ensuring that LLMs are not only powerful but also trustworthy and aligned with human values.

\noindent \textbf{Current State.} The current state of LLM testing and evaluation, while rapidly evolving, still falls short of a holistic approach, with practices often fragmented and heavily reliant on standardized benchmarks that may not fully capture real-world performance or ethical considerations~\cite{sun2024trustllm,chao2024jailbreakbench,chang2024survey,liu2024your,cao2024concerned}. This landscape can be broadly categorized into three critical dimensions: helpfulness, honesty, and harmlessness, each presenting unique challenges. Functionality and performance evaluation~(helpfulness) primarily relies on benchmark datasets for tasks such as question answering, task completion, and knowledge retrieval~\cite{sun2024scieval,guo2023can,kandpal2023large}, but often falls short in capturing real-world complexities, particularly in domains like code generation~\cite{liu2024your,du2024evaluating}. The issue of data contamination further complicates performance assessment~\cite{deng2024benchmark,magar2022data}. Reliability and truthfulness assessment~(honesty) focuses on factual correctness and consistency, employing benchmarks like TruthfulQA~\cite{lin2021truthfulqa} and FELM~\cite{zhao2024felm}, yet struggles with detecting subtle misinformation and balancing hallucination mitigation with generative capabilities~\cite{ji2023hallucination,zhang2023siren}. Safety and ethical compliance evaluation~(harmlessness) encompasses a wide range of evaluations for biases, security vulnerabilities, and potential misuse~\cite{gehman2020realtoxicityprompts,huang2023flames,zhao2024evaluating}, but faces challenges in capturing context-dependent ethical considerations and anticipating novel misuse scenarios~\cite{bhatt2023purple,zhang2023safety}. 

\noindent \textbf{Challenge I~[\underline{SE Perspective}]: Functionality and Performance Evaluation.}
Evaluating the helpfulness of LLMs is a critical aspect of ensuring their practical utility and widespread adoption. To this end, researchers have been developing benchmark datasets and tasks that measure LLM performance on capabilities such as question answering~\cite{sun2024scieval,bian2021benchmarking,zhao2024felm}, task completion~\cite{guo2023can,valmeekam2024planbench,efrat2022lmentry,dong2023codescore,li2023taco}, and knowledge retrieval~\cite{kandpal2023large,shi2024corecode,gu2024xiezhi} across diverse domains. These benchmarks not only test for general knowledge~\cite{kandpal2023large,shi2024corecode,gu2024xiezhi} but also probe domain-specific expertise~\cite{sun2024scieval,guo2023can,valmeekam2024planbench}, allowing for a comprehensive assessment of an LLM's ability to provide useful and relevant outputs. However, there are still several formidable challenges that highlight not only the complexity inherent in measuring the utility of such models but also underscore the necessity for ongoing refinement in our approaches to evaluation.

\begin{itemize}[leftmargin=15pt]
    \item \textbf{Opportunity: Developing Comprehensive Metrics \& Benchmarks.} First and foremost, the opportunity to develop more comprehensive metrics and benchmarks provides a pathway to better understand the performance of LLMs~\cite{liu2024your,yu2024codereval,mcintosh2024inadequacies}. Traditional benchmarks, while useful, often fail to capture the multifaceted nature of tasks LLMs are expected to perform, especially in areas like code generation~\cite{liu2024your,du2024evaluating,yu2024codereval}. While some repository-level datasets have been proposed to address this issue~\cite{wang2024repos,liu2023repobench}, they remain limited in scope and do not fully encompass the complexities involved in generating code at the project level. This highlights a significant gap in current benchmarking efforts, indicating a need for more robust metrics that can evaluate an LLM's ability to understand project-specific contexts, manage dependencies across multiple files, and ensure consistency within a larger codebase. Developing such metrics remains a critical area for research and innovation.
    % \item \textbf{Opportunity: Developing Comprehensive Metrics \& Benchmarks.} First and foremost, the opportunity to develop more comprehensive metrics and benchmarks provides a pathway to better understand the performance of LLMs~\cite{liu2024your,yu2024codereval,mcintosh2024inadequacies}. Traditional benchmarks, while useful, often fail to capture the multifaceted nature of tasks LLMs are expected to perform, especially in areas like code generation~\cite{liu2024your,du2024evaluating,yu2024codereval}.\yj{repo level has been proposed} The current benchmarks, such as HumanEval~\cite{chen2021evaluating} and AiXBench~\cite{hao2022aixbench}, provide a starting point but do not sufficiently address the complexities of generating code at the repository or project level~\cite{liu2024your,du2024evaluating,yu2024codereval}. This limitation points to a need for benchmarks that can assess an LLM's ability to understand project-specific contexts, manage dependencies across multiple files, and ensure consistency within a larger codebase. Developing such metrics requires a deep understanding of the practical tasks and a thoughtful consideration of how to measure success in those tasks.
    
    \item \textbf{Opportunity: Mitigating Data Contamination.} 
    The issue of data contamination~\cite{deng2024benchmark,magar2022data,cao2024concerned} significantly complicates the evaluation of LLMs. Data contamination occurs when a model is inadvertently exposed to information from the test set during training, leading to inflated performance metrics that do not accurately represent the model's true capabilities~\cite{li2024task,ponnuru2024unveiling}. This challenge is particularly acute in domains like code generation~\cite{dong2024generalization,cao2024concerned}, where the vast amount of publicly available code means that models might ``learn'' specific solutions during training that they later reproduce during testing. Such instances of data contamination not only overestimate the model's performance but also obscure our understanding of its ability to generate innovative solutions to new problems.
    While there have been efforts to quantify and detect data contamination~\cite{li2024latesteval,golchin2023time,li2023estimating,golchin2023data}, effectively addressing this issue remains a challenge~\cite{dekoninck2024evading,balloccu2024leak,oren2023proving,cao2024concerned}. Opportunities in identifying and mitigating the impact of data contamination include the development of novel evaluation frameworks that can detect when a model is reproducing rather than generating solutions~\cite{golchin2023time,li2023estimating,golchin2023data}. Moreover, there is a demand for advanced perturbation techniques~\cite{kargupta2003privacy} that preserve the integrity of the code while adding enough variability to prevent contamination. This involves developing flexible perturbation frameworks specifically designed for semantic class code evaluation datasets, capable of incorporating different approaches such as code refactoring and transformations that maintain logical equivalence. Additionally, developing testing metrics and benchmarks specifically designed to prevent data contamination will be crucial in ensuring reliable evaluations of LLM performance~\cite{dong2024generalization,jain2024livecodebench,li2024latesteval}. 

\end{itemize}

\noindent \textbf{Challenge II~[\underline{SE Perspective}]: Reliability and Truthfulness Assessment.}
As LLMs become increasingly influential in various domains, ensuring their honesty and truthfulness is paramount to building trust and preventing the spread of misinformation. Reliability and truthfulness assessment~\cite{lin2021truthfulqa,zhao2024felm,liang2022holistic,li2023halueval,lee2022factuality,muhlgay2023generating} for LLMs involves assessing whether the models can consistently provide information that is not only factually correct but also free from deception or misleading implications. These tests aim to identify instances of hallucinated~\cite{ji2023hallucination,zhang2023siren,agarwal2024codemirage} or fabricated information~\cite{lin2021truthfulqa,lee2022factuality,muhlgay2023generating} in LLM outputs, which can undermine their trustworthiness. Assessing the consistency and coherence of LLM outputs across multiple queries and prompts can reveal potential inconsistencies or contradictions, which may indicate a lack of factual grounding or honesty. 

\begin{itemize}[leftmargin=15pt]
    \item \textbf{Opportunity: Hallucination Mitigation.} Hallucination mitigation in LLMs is an area of significant concern, with various innovative techniques employed to address the issue~\cite{tonmoy2024comprehensive,gunjal2024detecting,alshahwan2024assure}. These methods range from retrieval augmented generation~\cite{gao2022rarr,peng2023check,varshney2023stitch} to self-refinement through feedback and reasoning~\cite{si2022prompting,ji2023towards}, each targeting different aspects of hallucination to ensure the accuracy and reliability of LLM outputs. However, there's an open problem in balancing mitigation efforts with the preservation of LLMs' generative capabilities, avoiding over-restriction that could stifle their performance. The development of LLMs with inherent mechanisms to prevent hallucinations is an exciting avenue, potentially leading to inherently more honest models.
\end{itemize}

\noindent \textbf{Challenge III~[\underline{S\&P + SE Perspective}]: Safety and Ethical Compliance Evaluation.}
The challenge of evaluating LLMs for harmful outputs and behaviors is multifaceted. Researchers have been developing benchmarks~\cite{gehman2020realtoxicityprompts,cui2024risk,huang2023flames,ji2024beavertails,nagireddy2024socialstigmaqa} that probe LLMs for the presence of harmful biases, stereotypes, or discriminatory language across various sensitive topics and demographic groups. 
Furthermore, testing LLMs for potential vulnerabilities to adversarial attacks~\cite{zhao2024evaluating,wang2023adversarial,qi2024visual},jailbreaks~\cite{zhou2024easyjailbreak,deng2024masterkey,chu2024comprehensive,liu2023jailbreaking,dong2024evaluating}, or misuse~\cite{bhatt2023purple,zhang2023safety,fu2023misusing,hazell2023spear,yao2024survey,madani2023metamorphic} by attackers ensures their outputs do not enable harmful actions or security breaches. 
Yet they are beset with the inherent challenge of predicting and counteracting the myriad ways in which these sophisticated models might be exploited or go awry.

\begin{itemize}[leftmargin=15pt]
    \item \textbf{Opportunity: Security Testing.} The domain of harmlessness testing for LLMs presents substantial opportunities to enhance the security and robustness of these systems. Developing sophisticated jailbreak prevention techniques~\cite{xu2022dense,zeng2024autodefense,yi2024jailbreak} can significantly improve LLM resilience against malicious attempts to bypass safety constraints. Furthermore, the implementation of comprehensive LLM red teaming protocols~\cite{ge2023mart,xu2024redagent,mazeika2024harmbench} offers a proactive approach to identifying and mitigating potential vulnerabilities before deployment. Such efforts can involve simulating various attack scenarios, from subtle prompt injections to more complex multi-turn manipulation attempts, thereby uncovering hidden weaknesses in LLM systems. Additionally, advancing adversarial testing methodologies~\cite{zhao2024evaluating,wang2023adversarial,zou2023universal} can lead to the development of more robust models capable of maintaining safe and ethical outputs even under targeted attacks.
    
    \item \textbf{Opportunity: Bias Mitigation.} Despite these challenges, the domain of harmlessness testing for LLMs presents substantial opportunities to enhance the safety and integrity of LLMs. Developing advanced benchmarks and testing protocols offers a pathway to not only detect but also rectify harmful outputs before they reach end-users, thereby safeguarding public trust in LLM applications. This endeavor encourages the creation of more nuanced and context-aware models, capable of discerning and adapting to the ethical implications of their outputs. Additionally, addressing the risks of adversarial misuse opens avenues for innovative defensive strategies, fortifying LLMs against manipulation and ensuring their outputs remain aligned with ethical standards. 
\end{itemize}

\subsection{Model: Release \& Sharing}
\label{sec:release}
The release and sharing phase represents a pivotal point in the LLM lifecycle, where trained models are packaged for distribution, complete with serialization and documentation detailing their capabilities, limitations, and intended applications.  These models are then published to repositories or model hubs like Hugging Face~\cite{huggingface}, making them accessible for reuse by others through techniques such as feature extraction, fine-tuning, transfer learning, and knowledge distillation~\cite{jiang2023empirical,taraghi2024deep,davis2023reusing}.

From an SE perspective, the vision for LLM release and sharing focuses on creating a seamless, efficient ecosystem for model distribution and reuse. This future envisions advanced platforms and protocols that facilitate easy access to models while ensuring version control, dependency tracking, and comprehensive documentation. The goal is to enable developers and researchers to build upon existing models efficiently, fostering innovation and rapid advancement in AI capabilities.

From an S\&P perspective, the vision emphasizes a robust framework for secure and responsible model sharing. This approach ensures that every shared model is accompanied by a detailed risk assessment, clear usage guidelines, and mechanisms to monitor and mitigate potential misuse. The vision includes the development of sophisticated tracking systems to monitor model propagation and adaptation, enabling swift responses to emerging security threats or ethical concerns throughout the LLM supply chain.

\noindent \textbf{Current State.} The current state of model release and sharing in the LLM supply chain, while evolving rapidly, still falls short of the envisioned ideal. Platforms like Hugging Face have significantly improved accessibility and collaboration in model sharing. However, the management of supply chain risks remains a critical challenge, particularly in terms of model provenance and security assurances~\cite{kathikar2023vulhug}. Model cards and associated documentation, while intended to provide comprehensive information, often fail to accurately reflect the true nature and capabilities of models~\cite{ajibode2024towards,pepe2024hugging,liang2024systematic,jiang2023empirical}. This weak model provenance leaves the ecosystem vulnerable to malicious model poisoning and other forms of tampering~\cite{zhao2024models,daniel2023poisongpt}.
The reuse and adaptation of models through techniques like fine-tuning introduce complex model dependencies that can lead to risk propagation. However, the open-source model ecosystem currently lacks robust frameworks for modeling and governing these interdependencies. Models are essentially treated as black boxes, and unlike traditional open-source software, static inspection offers limited security assurances. Vulnerable pre-trained models may harbor hidden biases, backdoors, or other malicious features that evade standard safety evaluations~\cite{abdali2024securing,zhao2024models,daniel2023poisongpt}.
License management for LLMs remains a contentious issue, as exemplified by the ongoing debates surrounding the licensing terms of models like Llama~\cite{charlie2023llamalicense,sriram2024llama3license}. The field is grappling with balancing open collaboration with intellectual property concerns and responsible AI development.
Furthermore, the security of model hubs themselves has come under scrutiny. Services like model conversion tools hosted on platforms such as Hugging Face have been shown to be vulnerable to manipulation~\cite{eoin2024hfconversion}, potentially allowing the introduction of malicious code into LLMs. The popularity of collaborative model merging techniques, while fostering innovation, also presents opportunities for introducing vulnerabilities~\cite{hammoud2024modelmerging}.

% Providing licensing information and metadata is crucial for facilitating responsible adoption and collaboration.  However, akin to traditional software supply chains, the reuse of pre-trained models introduces significant supply chain risks that must be carefully managed. The propagation of dependency risks, such as privacy concerns~\cite{yang2024robustness,abdali2024securing}, biases~\cite{hutiri2023tiny,brown2022does,ferrara2023should}, hallucinations~\cite{liu2024survey,zhang2023siren}, and vulnerabilities~\cite{abdali2024securing}, can occur throughout the supply chain during model reuse and adaptation processes.  Ensuring the trustworthy and responsible use of these powerful models necessitates comprehensive supply chain risk management strategies to mitigate potential threats and foster transparency, compliance, and accountability.

\noindent \textbf{Challenge I~[\underline{S\&P + SE Perspective}]: Model Dependency Analysis.} 
While model cards and associated documentation provide initial insights into model characteristics, they fall short in offering comprehensive dependency information for LLMs~\cite{ajibode2024towards}. The challenge of conducting thorough model dependency analysis for LLMs extends far beyond these basic documentation practices. Existing dependency analysis approaches, primarily developed for traditional deep learning models~\cite{li2021modeldiff,klabunde2023similarity}, include training-code analysis, model file analysis, and watermarking techniques. However, these methods face significant limitations when applied to LLMs. A key challenge for training-code analysis is that the training code for some LLMs is not open-source, making this method infeasible in those cases. Binary model file analysis, while potentially precise, is extremely resource-intensive for LLMs given their size. Watermarking techniques, while useful for establishing model ownership and provenance, have limited utility in constructing comprehensive dependency relationships, especially for third-party analysts who may struggle to detect or interpret these watermarks accurately. The unique characteristics of LLMs, including their massive scale, complex training processes, and often opaque development pipelines, pose unprecedented challenges to traditional dependency analysis methods. Consequently, developing robust and efficient dependency analysis techniques specifically tailored to LLMs remains an open problem.

\begin{itemize}[leftmargin=15pt]
    \item \textbf{Opportunity: Comprehensive Strategies for Model Dependency Tracking in LLMs.} Addressing the challenges of dependency tracking in LLMs presents several opportunities for advancing the field. Developing more comprehensive and standardized documentation practices beyond model cards could provide deeper insights into LLM dependencies and provenance. For scenarios where training code is not available, research could focus on creating novel inference techniques to deduce potential dependencies from model behavior or output. To address the resource-intensive nature of binary model file analysis for LLMs, opportunities exist to develop more efficient algorithms or sampling techniques that can provide accurate dependency information without analyzing the entire model. Creating tools for dynamic tracking of dependencies during model inference could offer real-time insights into how different components interact in practice. Investigating ways to integrate dependency analysis with existing security frameworks could enhance the robustness of LLMs against supply chain attacks.
\end{itemize}

\noindent \textbf{Challenge II~[\underline{S\&P + SE Perspective}]: Risk Propagation in LLM Supply Chains.} 
A significant challenge in the LLM supply chain lies in quantifying the propagation of risks across different models and components. The complex nature of LLMs, with their intricate architectures and numerous dependencies, makes it difficult to measure how vulnerabilities such as privacy breaches, biases, hallucinations, or potential backdoors propagate from one model to another. This challenge is exacerbated by the lack of standardized metrics and methodologies for assessing risk transmission.
Model reuse techniques~\cite{jiang2023empirical}, including fine-tuning, transfer learning, distillation, and model merging, add further complexity to this issue. When models are reused, there is a risk that vulnerabilities inherent in the original model can be unintentionally transferred to new iterations or applications. For instance, if a base model contains biases, these biases may be amplified or altered during the distillation process, leading to unforeseen consequences in downstream applications. Quantifying how such biases propagate through different stages of model reuse remains an open problem.
Similarly, measuring the extent to which privacy vulnerabilities in training data persist through model iterations and affect derived models is not straightforward. The challenge is not only to identify these vulnerabilities but also to understand how they evolve as models are reused. 

% \yj{model reuse, model distillation}
% A significant challenge in the LLM supply chain lies in quantifying the propagation of risks across different models and components. The complex nature of LLMs, with their intricate architectures and numerous dependencies, makes it difficult to measure how vulnerabilities such as privacy breaches, biases, hallucinations, or potential backdoors propagate from one model to another. This challenge is exacerbated by the lack of standardized metrics and methodologies for assessing risk transmission. For instance, quantifying how a bias present in a base model affects fine-tuned versions or downstream applications remains an open problem. Similarly, measuring the extent to which privacy vulnerabilities in training data persist through model iterations and affect derived models is not straightforward. The absence of clear propagation metrics also hinders the ability to assess the effectiveness of mitigation strategies. 

\begin{itemize}[leftmargin=15pt]
    \item \textbf{Opportunity: Developing Model Bill of Materials~(MBOM).} There lies a significant opportunity to enhance the security and integrity of LLMs through the development of standardized practices for generating and maintaining a MBOM for pre-trained models~\cite{david2024mbom}, mirroring the concept of SBOM. Such standardization would improve supply chain transparency, enabling stakeholders to more effectively identify, assess, and mitigate risks. Moreover, fostering collaboration among researchers, industry practitioners, and regulatory bodies can lead to the establishment of robust best practices and guidelines for the responsible release and sharing of models. This collaborative approach would not only enhance the trustworthiness and accountability of LLMs across the supply chain but also ensure that risk mitigation strategies are holistic, timely, and aligned with evolving ethical and security standards, ultimately leading to a safer and more reliable LLM ecosystem.
\end{itemize}

\noindent \textbf{Challenge III~[\underline{SE Perspective}]: Open-Source Model License Compatibility.}
The increasing prevalence and complexity of open-source LLMs present significant challenges in terms of license compatibility and management. The challenge lies in accurately identifying and resolving potential conflicts between different open-source licenses, such as copyleft vs. permissive licenses, or addressing compatibility issues between commercial use restrictions. This complexity is further compounded when open-source models are used as foundations for downstream applications or fine-tuned models, potentially introducing additional licensing considerations. Moreover, the dynamic nature of open-source development, with frequent updates and contributions from diverse sources, makes maintaining up-to-date license compliance a continuous challenge.

\begin{itemize}[leftmargin=15pt]
    \item \textbf{Opportunity: Automated License Analysis.} Developing automated license analysis frameworks specifically designed for pre-trained models could significantly improve compliance management, drawing on techniques from software license management and natural language processing to enhance accuracy and efficiency. Furthermore, integrating license compliance checks directly into model development and deployment pipelines offers a proactive strategy to ensure ongoing adherence to licensing requirements, potentially establishing new standards for license management practices. In addition, creating a robust benchmark dataset for open-source licenses will be crucial for training and evaluating these automated systems. This dataset could aid in identifying license terms and conditions while also providing examples of conflicts, thereby enhancing the reliability of automated analyses.
\end{itemize}

\noindent \textbf{Challenge IV~[\underline{S\&P Perspective}]: Pre-trained Model Code Poisoning/Backdoor.}
The proliferation of model hubs as centralized platforms for sharing and distributing pre-trained models introduces significant security challenges. These hubs, while fostering collaboration and accessibility, also become potential points of failure and attractive targets for malicious actors. Key challenges include ensuring the integrity and authenticity of uploaded models, preventing the distribution of models with embedded malware or backdoors~\cite{zhao2024models}, and protecting against unauthorized access or data breaches. The dynamic nature of these platforms, with frequent updates and contributions from various sources, further complicates maintaining consistent security standards. Moreover, the potential for supply chain attacks, where compromised models could affect numerous downstream applications and users, presents a systemic risk to the LLM ecosystem.

\begin{itemize}[leftmargin=15pt]
\item \textbf{Opportunity: Advanced Security Solutions for Model Hub Integrity.}
The security challenges posed by model hubs open up significant opportunities for developing robust security frameworks and technologies. There is potential to create advanced model verification systems that can automatically scan for malware, backdoors, or other vulnerabilities in uploaded models~\cite{zhao2024models}. Opportunities exist to implement measures to reduce model poisoning attack vectors. This could include developing sophisticated typo-squatting detection mechanisms to prevent malicious actors from exploiting similar or deregistered model names, organizations, or namespaces, addressing both typo-squatting and AIJacking vulnerabilities~\cite{adrian2023modelconfusion,legit2023aijacking}. Enhanced token management systems could mitigate the risk of API token leaks from prominent organizations~\cite{lasso2023token}, thereby preventing unauthorized access and potential supply chain attacks. These advancements would not only enhance the security of individual model hubs but also contribute to building a more resilient and trustworthy LLM development ecosystem, reducing the risk of widespread supply chain attacks and unauthorized model access or manipulation~\cite{eoin2024hfconversion}.
\end{itemize}

\subsection{Model: Deploy \& Maintenance}
\label{sec:deploy}
In the rapidly evolving landscape, pre-trained models must adapt to changing real-world conditions, emerging data distributions, and novel task requirements to maintain their utility and relevance. The model maintenance and update phase is crucial for ensuring the longevity and continued effectiveness of these powerful models. However, this phase presents several opportunities and challenges that demand rigorous exploration by the research community.

\noindent \textbf{Challenge I~[\underline{SE Perspective}]: Model Drift.}
The challenge of identifying and quantifying model drift in LLMs is considerable~\cite{madaan2023detail,li2024measuring}. LLMs are trained on vast datasets that are supposed to represent the linguistic diversity of their intended application domain. However, as the language evolves or the model is applied to slightly different contexts, the ability to remain relevant and consistent can shift in subtle ways. Recent research~\cite{madaan2023detail,broscheit2022distributionally} emphasizes the need for sophisticated tools that can detect not only overt drifts in language usage but also more nuanced shifts in sentiment, context, or cultural references. These tools must be capable of parsing the complexities of human language, requiring ongoing refinement and adaptation to new linguistic phenomena.

\begin{itemize}[leftmargin=15pt]
    \item \textbf{Opportunity: Model Drift Monitoring.} The realm of drift monitoring in LLMs presents a fertile ground for innovation and development. There is a significant opportunity to create and refine tools that can accurately detect and measure drift in various dimensions, from language usage to sentiment and contextual nuances. Furthermore, integrating these drift monitoring tools into the model development and deployment lifecycle can provide ongoing insights into model performance~\cite{madaan2023detail}, enabling timely adjustments and enhancements. This proactive approach to managing model drift not only ensures the sustained relevance and accuracy of LLMs but also opens new avenues for research in understanding and mitigating the subtleties of language evolution in artificial intelligence.
\end{itemize}

\noindent\textbf{Challenge II~[\underline{SE Perspective}]: User Feedback Incorporation.} 
Effectively incorporating user feedback into the maintenance and updating of LLMs presents a significant challenge. While user feedback can provide valuable insights into model performance and areas for improvement, the process of collecting, analyzing, and integrating this feedback into the model lifecycle is complex~\cite{shi2024wildfeedbackaligningllmsinsitu}. Users may have diverse perspectives and expectations, leading to varied feedback that can be difficult to standardize. Additionally, there is a risk that incorporating feedback without proper validation could introduce biases or unintended consequences, further complicating the model's reliability and effectiveness.

\begin{itemize}[leftmargin=15pt]
\item \textbf{Opportunity: Developing User Feedback Systems.}  Developing robust systems that facilitate the collection and analysis of user feedback can enhance the understanding of model performance in real-world applications. These systems could leverage techniques such as active learning and reinforcement learning to prioritize and integrate user input effectively. Furthermore, creating intuitive interfaces for users to provide feedback will encourage participation and ensure that diverse perspectives are captured. By systematically incorporating validated user feedback into the model development lifecycle, researchers can improve model accuracy, relevance, and user satisfaction while also fostering a more collaborative relationship between users and LLM systems. 
\end{itemize} 
\section{Downstream Ecosystem}
\label{sec:ecosystem}
The downstream application ecosystem serves as the final stage in the LLM supply chain, embodying the point where the efforts invested in developing, training, and refining these models are translated into practical benefits across different fields. This ecosystem is characterized by a diverse array of applications and services that leverage pre-trained models to address real-world challenges, driving innovation and efficiency.

\subsection{Application: LLM Chatbots and Applications}
\label{sec:chatbot}
LLM-powered chatbots and applications (apps) represent a significant advancement in the downstream ecosystem of the LLM supply chain. These applications leverage the capabilities of LLMs to provide interactive, intelligent, and task-specific solutions across various domains. From customer service chatbots to specialized productivity tools, LLM apps are transforming how users interact with AI technology. 

\noindent \textbf{Current State.} Platforms like the GPT Store~\cite{gptstore} are emerging as centralized hubs where developers can publish their LLM-powered apps (i.e., GPTs), and users can discover and access these tools to fulfill a wide array of tasks and objectives. This ecosystem draws parallels with mobile app stores~\cite{fu2013people,lee2014determinants,martin2016survey}, aiming to create a curated, secure, and user-friendly environment for LLM-driven apps. The proliferation of LLM chatbots and apps is catalyzing innovation by lowering the barriers to entry for developers and providing them with platforms to build and deploy LLM-powered solutions. However, this rapid growth and accessibility also introduce new vulnerabilities and governance challenges. As these platforms evolve, they must grapple with novel security threats specific to LLM applications~\cite{hou2024security,su2024gpt}, such as prompt injection attacks and the potential misuse of advanced features like function calling. Furthermore, the unique nature of LLM apps, which can generate and manipulate content in real time, presents unprecedented challenges in terms of quality control, ethical considerations, and regulatory compliance.

\noindent \textbf{Challenge I~[\underline{S\&P Perspective}]: Security Vulnerabilities in LLM Apps.}
LLM-powered apps face significant security challenges, particularly in the realms of prompt injection attacks, misuse of function calling features, and prompt leakage~\cite{wu2024dark,owasp2024llmv2}. Prompt injection attacks can manipulate the behavior of LLM chatbots, potentially leading to unauthorized actions or the disclosure of sensitive information~\cite{liu2023prompt,greshake2023not,liu2023promptbench}. Function calling features, while powerful, introduce new attack vectors. Attackers may exploit these features to extract parameters and sensitive information about the underlying tools, hijack data flows, or conductresource exhaustion attacks~\cite{owasp2024llmv2} by overloading the system with malicious function calls~\cite{owasp2024llmv2}. Additionally, prompt leakage poses a significant risk, where sensitive information embedded in prompts may be inadvertently revealed through the model's responses, potentially compromising user privacy and system security~\cite{hou2024security,su2024gpt}.

\begin{itemize}[leftmargin=15pt]
\item \textbf{Opportunity: Advanced Security Measures for LLM Apps.} 
The security challenges present opportunities for developing robust defense mechanisms. There is potential for creating advanced prompt filtering and sanitization techniques to mitigate prompt injection risks. Implementing strict input validation and output sanitization for function calls could significantly reduce the risk of malicious exploitation. Developing intelligent monitoring systems that can detect and prevent abnormal patterns of function usage could help in identifying and mitigating potential attacks in real time. Furthermore, there's an opportunity to establish industry-wide security standards and best practices specifically tailored for LLM-powered applications, fostering a more secure ecosystem.
\end{itemize}

\noindent \textbf{Challenge II~[\underline{SE + S\&P Perspective}]: App Store Governance.} 
The creation and management of an LLM app store present significant challenges in terms of quality control, compatibility, and ethical considerations~\cite{zhao2024llm,su2024gpt}. Ensuring that each LLM-powered application adheres to high standards of accuracy, fairness, and security is crucial for maintaining user trust and compliance with regulatory standards. Security and ethical concerns are at the forefront, as the store must implement stringent policies to prevent the dissemination of models that could be used maliciously or propagate bias, misinformation, or harmful content~\cite{hou2024security}. In addition, LLM app stores are required to implement strong safeguards to protect children from inappropriate content and ensure that apps are designed with children’s best interests in mind.

\begin{itemize}[leftmargin=15pt]
\item \textbf{Opportunity: App Store Measurement.}
The challenges in LLM app store governance open up several opportunities for innovation in security and management systems. Firstly, there is significant potential in leveraging user reviews and feedback for research purposes. By analyzing user comments and ratings, researchers can gain valuable insights into the performance, usability, and potential issues of LLM apps~\cite{hou2024voices,su2024gpt}, thereby informing future developments and improvements in the field. Secondly, the need for quality control presents an opportunity to develop sophisticated malicious LLM app detection systems. These systems could employ advanced techniques to identify apps that may pose security risks, exhibit unethical behavior, or violate platform policies~\cite{hou2024security}. Lastly, the complex regulatory landscape offers an opportunity to create automated compliance checking tools, which could help developers ensure their LLM apps meet the necessary legal and ethical standards across different jurisdictions, streamlining the app submission and approval process.
\end{itemize}

% Creating an LLM app store introduces several challenges, primarily concerning the quality control, compatibility, and ethical considerations of the models hosted~\cite{zhao2024llm,su2024gpt}. Ensuring that each LLM adheres to a high standard of accuracy, fairness, and security is crucial to maintaining user trust and compliance with regulatory standards. Additionally, the diversity of LLMs in terms of size, functionality, and intended use cases necessitates robust mechanisms for assessing and certifying model compatibility with various platforms and user requirements. Ethical concerns also come to the forefront, as the store must have stringent policies to prevent the dissemination of models that could be used maliciously or propagate bias, misinformation, or harmful content~\cite{hou2024security}. However, an LLM app store also presents vast opportunities for innovation and value creation. By implementing mechanisms for user engagement, such as ratings and reviews~\cite{hou2024voices}, the store can facilitate a feedback loop that drives the evolution of more sophisticated and user-aligned LLMs, promoting a culture of transparency and accountability within the LLM community.

\subsection{Application: Autonomous LLM-based Agents}
\label{sec:agent}
Autonomous LLM-based agents~(ALAs) represent a cutting-edge application of LLMs, offering autonomous or semi-autonomous task execution across various domains. These agents leverage the advanced reasoning and knowledge synthesis capabilities of LLMs to perform complex tasks, make decisions, and interact with users and systems in sophisticated ways. From virtual assistants and automated workflow managers to more specialized agents in fields like finance, healthcare, and research, ALAs are pushing the boundaries of LLM application~\cite{talebirad2023multi}. The development of these agents is facilitated by frameworks and platforms that allow for the integration of LLMs with external tools and data sources, enabling more powerful and versatile LLM systems~\cite{mei2024llm}. 

\noindent \textbf{Current State.}
The current landscape of ALAs is characterized by rapid advancement and diverse applications across multiple domains. Recent developments have seen the emergence of more sophisticated agent architectures that combine LLMs with external tools, knowledge bases, and decision-making frameworks~\cite{guo2024large,xi2023rise}. These agents are increasingly capable of performing complex, multi-step tasks with minimal human intervention. For instance, in the field of software development, ALAs are being used for code generation, debugging, and even system design~\cite{jin2024llms,liu2024agents,huang2024agents}. However, alongside these advancements, there is growing concern about the ethical implications and potential risks associated with increasingly autonomous agents~\cite{he2024emerged,li2024personal,zhang2024breaking}. 
% Current research is focused on enhancing the reliability, transparency, and controllability of ALAs, with particular emphasis on developing robust safety measures and alignment techniques to ensure agent behaviors remain consistent with human intentions and values. 

\noindent \textbf{Challenge I~[\underline{SE Perspective}]: Agent Optimization.}
From a software engineering perspective, a significant challenge in ALA development is the optimization of agent performance through efficient workflow design and orchestration~\cite{sun2024llm}. This challenge falls under the emerging field of Software Engineering for LLM (SE4LLM). The complexity of tasks that ALAs can handle requires careful consideration of how different components interact, how information flows between stages of processing, and how to balance autonomy with controlled behavior. Effective agent optimization involves designing modular and reusable components, implementing efficient error handling and recovery mechanisms, and creating adaptive workflows that can adjust to varying task requirements and environmental conditions~\cite{liu2024agents,huang2024agents}.

\begin{itemize}[leftmargin=15pt]
\item \textbf{Opportunity: Workflow Orchestration Frameworks.}
This challenge presents an opportunity to develop sophisticated workflow orchestration frameworks specifically tailored for ALAs. Such frameworks could provide tools for visual workflow design, automated performance analysis, and dynamic workflow adjustment based on real-time feedback. Research in this area could focus on creating best practices for agent workflow design~\cite{liu2024agents}, developing metrics for agent efficiency and effectiveness, and exploring ways to optimize the interaction between LLM components and external tools or data sources.
\end{itemize}

\noindent \textbf{Challenge II~[\underline{S\&P Perspective}]: Vulnerable Agent Logic.}
Another critical security challenge in ALA systems is the presence of vulnerable agent logic~\cite{owasp2024llmv2}. ALAs rely on non-deterministic outcomes and inherit the overreliance and safety issues inherent to LLMs, but with potentially greater cascading effects. Validating agent behavior is problematic and could be non-exhaustive, potentially leading to catastrophic unintended consequences. Additionally, adversaries can identify and exploit gaps in agent logic to achieve malicious outcomes. The complexity and opacity of LLM decision-making processes make it difficult to anticipate all possible failure modes or attack vectors, creating significant security risks in deployed agent systems~\cite{zhang2024breaking,he2024emerged,wang2024agents}.

\begin{itemize}[leftmargin=15pt]
\item \textbf{Opportunity: Robust Agent Logic Validation.}
This challenge opens up opportunities for developing advanced techniques in agent logic validation. Research could focus on creating comprehensive testing frameworks that combine formal verification methods with adversarial testing approaches. There's potential for developing tools that can automatically analyze agent logic for vulnerabilities, simulate various attack scenarios, and provide actionable insights for improving agent robustness. Additionally, research into explainable LLM techniques specifically tailored for ALAs could help in better understanding and validating agent decision-making processes, thereby enhancing security and trustworthiness.
\end{itemize}

\noindent \textbf{Challenge III~[\underline{S\&P Perspective}]: Agent Autonomy Escalation.}
A significant security concern in the deployment of ALAs is the risk of autonomy escalation~\cite{owasp2024llmv2}. This occurs when an LLM-based agent gains unintended levels of control or decision-making capabilities, potentially leading to harmful or unauthorized actions. Such escalation can result from misconfigurations, unexpected interactions between different agents, or exploitation by malicious actors. The complexity of ALAs, combined with their ability to learn and adapt, makes it challenging to predict and control their behavior in all possible scenarios. This unpredictability poses risks to system integrity, data security, and user safety. Moreover, as agents become more integrated into critical systems and workflows, the potential impact of autonomy escalation increases.

\begin{itemize}[leftmargin=15pt]
\item \textbf{Opportunity: Agent Permission Mechanism.}
The challenges posed by agent autonomy escalation present opportunities for crucial security research focused on real-world ALA systems. A primary area of investigation should be the analysis and improvement of permission mechanisms in existing agent deployments. Researchers can study how current systems implement and enforce access controls, identifying potential vulnerabilities that could lead to unauthorized escalation of privileges. There's an opportunity to develop more robust and granular permission models specifically designed for the dynamic nature of ALAs, incorporating principles of least privilege and zero trust architectures~\cite{tidjon2022never}.
\end{itemize}

\section{Conclusion}
\label{sec:conclusion}
This paper presents the first comprehensive research agenda for the LLM supply chain, identifying the complex challenges and opportunities arising from the rapid advancement of large language models. By examining the LLM supply chain through the dual lenses of software engineering and security \& privacy, we have provided a structured approach to understanding and advancing this critical field. Our analysis spans the infrastructure layer, foundation model layer, and downstream application ecosystem, presenting visions for robust and secure LLM development, reviewing current practices, and identifying key challenges and research opportunities. This work bridges the existing research gap in systematically understanding the multifaceted issues within the LLM supply chain. The proposed research agenda offers valuable insights to guide future efforts, fostering innovation while ensuring the responsible advancement of LLM technologies.

\section*{Acknowledgement}
This work was supported in part by the Key R\&D Program of Hubei Province (2023BAB017, 2023BAB079), the National Natural Science Foundation of China (grants No.62072046, 62302181), HUST CSE-HongXin Joint Institute for Cyber
Security, HUST CSE-FiberHome Joint Institute for Cyber Security, and the Xiaomi Young Talents Program.

\bibliographystyle{ACM-Reference-Format}
\bibliography{mainfix}

\end{document}